\documentclass[12pt]{article}
\usepackage{array,amsmath,amssymb,verbatim}
\usepackage{graphicx}
\usepackage{bm}
\csname @addtoreset\endcsname{equation}{section}

\newcommand{\ket}[1]{\left| #1\right\rangle}               

\newcommand{\braket}[2]{\left\langle #1 \right| \left.\! #2 \right\rangle}

\newcommand{\fbraket}[2]{\langle #1  |\! #2 \rangle}
\newcommand{\bra}[1]{\left\langle #1\right|}
\newcommand{\G}{\mathcal{G}}
\newcommand{\s}{\mathcal{S}}
\newcommand{\F}{\mathcal{F}}

\begin{document}

{\hbox to\hsize{April 2005 \hfill
{Bicocca--FT--05--7}}}

\begin{center}
\vglue .06in
{\Large \bf The renormalized and Renormalization--Group\\[0.3cm]
invariant Hartree--Fock approximation}
\\[.45in]
Claudio Destri\footnote{claudio.destri@mib.infn.it} ~and~
Andrea Sartirana\footnote{andrea.sartirana@mib.infn.it}\\
{\it Dipartimento di Fisica dell'Universit\`a degli studi di
Milano-Bicocca,\\ 
and INFN, Sezione di Milano, piazza della Scienza 3, I-20126 Milano,
Italy}\\[.8in]

{\bf ABSTRACT}\\[.0015in]
\end{center}

We study the renormalization problem for the Hartree--Fock approximation of
the $O(N)-$invariant $\phi^4$ model in the symmetric phase and show how to
systematically improve the corresponding diagrammatic resummation to
achieve the correct renormalization properties of the effective field
equations, including Renormalization--Group invariance with the one--loop
beta function. These new Hartree--Fock dynamics is still of mean field
type but includes memory effects which are generically nonlocal also in space.




\vskip 1truecm
\section{Introduction, summary and outlook}

Cosmology, particle and condensed matter physics have given, in recent
years, a great impulse to the search of a deeper qualitative and
quantitative understanding of out--of--equilibrium dynamics of quantum
fields. In fact a treatment based on first principle of the
late time and strongly coupled evolution of quantum systems would provide a
better insight in an important class of phenomena such as the reheating of
the universe after inflation or thermalization of the quark gluon plasma in
the ultra--relativistic heavy--ion and hadron colliders (RHIC, LHC).

The challenge in treating quantum field theories in non--equilibrium
conditions is that, except for very short time, standard perturbation
theory does not provide satisfactory results (this is true also at
equilibrium with nonzero temperature). Therefore one has to look for
nonperturbative approaches providing infinite partial resummations of
Feynman diagrams \cite{Berges:intro,Manfredini:2000sk}. The simplest of
such schemes are the mean field approximations such as the leading--order
large--N expansion
\cite{Cooper:largeN,Boyanovsky:1994me,Baacke:largeN,Destri:largeN} and the
Hartree, or Hartree--Fock (HF) variational method
\cite{Amelino-Camelia:1992nc,Destri:1999he,Michalski,Matsui}.

These approximations have been extensively studied and their features are
well known \cite{Berges:2000ur,Habib:irrev,Boyanovsky:meanfield,
  Wetterich:meanfield,Smit:meanfield,Baacke:2001zt}: they do provide a
backreaction term on the evolution of quantum fluctuations that stabilize
dynamics at late time but, on the other hand, they fail to properly
describe an important aspect of late time dynamics such as thermalization.
More elaborate approaches going beyond mean field have been put forward by
considering the 2PI (or 2PPI) effective action
\cite{Calzetta:1986cq,Cornwall:1974vz} at two (or more) loop or at
next--to--leading order in $1/N$ expansion
\cite{Berges:NLO,Cooper:2005vw,Baacke:BeyondHF}, yielding indeed
approximate numerical thermalization at strong coupling.

Apart from the search for a better description of the late-time
dynamics, there has recently been much progress in understanding more
formal aspects of resummed approximations such as their
renormalization properties. This is a long standing problem
\cite{Lenaghan:1999si,Pi:1987df} and recently a systematic method has
been developed \cite{Cooper:2004rs,Jakovac,Blaizot,vanHees} to remove
divergences in the $\Phi$--derivable approximations based on applying
a BPHZ subtraction procedure to diagrams with resummed propagators.

In this article we will consider the simple HF approximation of the $O(N)$
$\phi^4$ model as defined by variational principles, or equivalently by
resummation of daisy (or bubble) diagrams. It is known
\cite{Destri:1999he,Lenaghan:1999si,Pi:1987df} that the usual
renormalization of bare coupling and mass is not consistent, so that the
simple Hartree approximation is not really renormalizable. We will show
that this nonrenormalizability is due to the absence of leading
logarithmically divergent contributions coming from diagrams which are not
present in the standard HF resummation. We shall show that including also
these contribution, plus suitably chosen finite parts, yields indeed a
renormalized and Renormalization--Group invariant version of HF equations.
We point out that these improved equations, although still of mean field
type, are nonlocal in time (that is, there is memory) and generically
nonlocal also in space space, unlike the original ones.  We also verify
that all these nonlocalities disappears in the $N\to\infty$ limit, which is
well known to provide a renormalizable, RG invariant and local
approximation to the out--of--equilibrium dynamics of the model.

Section \ref{Generalities} is dedicated to generalities. In
subsec.~\ref{varHF} the HF approach is introduced as a variational
approximation with a Gaussian state functional in the Schr\"odinger
picture. The corresponding well known equations are rederived [see
eqs.~(\ref{HFeqskernels}) and eq.~(\ref{HFeqsmf})]. In subsec.~\ref{CTP} we
review the CTP formulation of out--of--equilibrium problems and the
equivalent definition of HF approximation as resummation of bubble
diagrams. In subsec.~\ref{PRCTP} we establish some relations between the
two approaches and define a generating functional $\F'$ [see
eqs.~(\ref{HFEOMgeneralform}) and~(\ref{Gammaformula})] whose proper
renormalization would render finite the dynamical problem.

In Section \ref{Renormalization} we tackle the central problem of
constructing renormalizable and RG invariant HF--like equations. In
subsec.~\ref{effpot} we begin by studying, as a leading example, the static
and homogeneous problem (i.e. the calculation of the effective potential)
for the $N=1$ case of a single scalar field. As already stated above, our
analysis leads naturally to add contributions from diagrams absent in the
HF approximation, in such a way to include all leading logarithmic
divergences and construct a modified effective potential [see
eqs.~(\ref{Veffmod})] with the correct renormalization properties. In
subsec.~\ref{general} the analysis of the HF approximation and its
renormalizable improvement are generalized to the fully dynamical and
inhomogeneous $N>1$ case [see. eqs.~(\ref{finaleqs})].  Further remarks on
the freedom of choosing various initial conditions without spoiling
renormalization and RG--invariance are made in subsec.~\ref{initstate}.

There are several possible developments along the lines of this work. First
of all it would be interesting to study numerically our modified HF field
equations, to investigate how the space-time nonlocalities affect the time
evolution as compared to the standard HF approximation, which is known to
fail even qualitatively at late times. Secondly, with the proper changes in
the renormalization scheme, the derivation could be extended to the case of
broken $O(N)$ symmetry, where the standard HF approximation faces other
infrared difficulties connected with the Goldstone theorem
\cite{Ivanov:2005yj}.  Since our improved HF resummation adds also
cutoff--independent contributions to the conventional one, it is
conceivable that infrared properties of the model are affected too. Another
challenging task is the extension of our approach to the full two--loop 2PI
effective action, through the inclusion of the nonlocal sunset diagram
which is absent by definition in any mean field approximation. In fact one
should expect that, even if such an inclusion allows to recover
renormalizability as compared to the conventional HF approximation, the
two--loop 2PI self-consistent equations still lack RG invariance with the
two--loop beta function, since the 2PI effective action does not contain
all diagrams which contribute to the next--to--leading ultraviolet
divergences.


\section{Generalities}\label{Generalities}

A non--equilibrium approach to quantum field theory is needed every time we have
to deal with an initial value problem. In fact, the usual in--out formalism of
QFT provides the means to calculate scattering expectation values
\begin{equation*}
  \langle in | \mathcal{O} ( t_1,t_2,\dots ) | out \rangle
\end{equation*}
where $\ket{in}$ and $\ket{out}$ represent particle states prepared at a
distant past and future, respectively, while $\mathcal{O}$ is some
observable depending on intermediate times $t_1,t_2,\dots$. On the other
hand, in order to study the real time dynamics from a given initial
condition we need to know amplitudes like
\begin{equation*}
   \bra{\Psi(t_0)}| \mathcal{O} ( t_1,t_2,\dots) \ket{\Psi(t_0)}
\end{equation*}   
where $\ket{\Psi}$ is a generic state prepared at a initial time $t_0
< t_1, t_2, \dots$. Out--of--equilibrium QFT provides the general setup for the
calculation of such matrix elements, as well as more general expectation
values in statistical mixtures of pure states like $\ket{\Psi}$.

In this section we briefly review some generalities about non--equilibrium
QFT and point out some properties that will be useful to derive the results
of the next section.


\subsection{Variational approach and HF approximation}\label{varHF}

A simple and intuitive approach to initial value problem is to treat QFT as
ordinary Quantum Mechanics of extended systems. We
restrict here to the case of interest for this paper: scalar field theory
in $3+1$ dimensions with quartic interaction and unbroken $O(N)$ symmetry.
The field variables are $\varphi_i ( \bm{x} )$ where $\bm{x}= (x_1,x_2,x_3)$
are space coordinates and $i=1,\dots,N $ is the $O(N)$ index. Classical
dynamics is defined by an action, functional of trajectories $\varphi_i ( x
)$, with $x= ( \bm{x}, t )$, in the form
\begin{equation}\label{classaction}
  S[\varphi]= \int d^4 x \left\{ \frac{1}{2} \partial_\mu 
  \varphi_i ( x ) \partial^\mu \varphi_i ( x ) - \frac{m^2}{2} 
  \varphi_i ( x ) \varphi_i ( x ) - \frac{\lambda}{4!} [\varphi_i ( x ) 
  \varphi_i ( x ) ]^2 \right\}
\end{equation}  
where $m^2$ and $\lambda$ are the squared mass and the coupling 
constant respectively. In terms of $\varphi_i ( \bm{x} )$ and its
canonical conjugated momentum $\pi_i (\bm{x})=\dot{\varphi}_i( \bm{x})$
the Hamiltonian reads
\begin{equation*} 
  H[\varphi ,\pi] = \int d^3 x \left[\frac{1}{2}\pi_i(\bm{x})\pi_i(\bm{x}) 
    + \frac{1}{2} \nabla \varphi_i (\bm{x})\cdot\nabla\varphi_i(\bm{x})
    + V ( \varphi_i ( \bm{x} ) ) \right]
\end{equation*}
where 
\begin{equation*}
  V(\varphi) = \frac{m^2}{2} \varphi_i \varphi_i + \frac{\lambda}{4!}
  ( \varphi_i \varphi_i )^2
\end{equation*}
Canonical quantization proceeds by imposing the standard commutation rules
\begin{equation*}
  \left[ \varphi_j ( \bm{x} ), \pi_k( \bm{y} )  \right] 
  = i \delta_{jk} \delta^3 ( \bm{x} - \bm{y} )
\end{equation*}
In the ``position'' representation states are functionals of 
$\varphi_i (\bm{x} )$ and the momentum operator reads
\begin{equation*}
  \pi_i( \bm{x} ) = -i \frac{\delta~~~}{\delta\varphi_i ( \bm{x} )}
\end{equation*}
It is then straightforward to write down the Hamiltonian operator and the 
corresponding Schr\"odinger equation.

In this framework, the standard variational methods of Quantum Mechanics
suggest an easy nonperturbative approximation to dynamics (see {\em e.g.}
\cite{Matsui,Eboli:1987fm,Cooper:1986wv}) . The
variational method follows from the observation that Schr\"odinger
equation is obtained by minimizing the Dirac action
\begin{equation*}
  S_{\rm D}[\Psi] = \int dt ~\bra{\Psi(t)}(i\partial_t-H)\ket{\Psi(t)}
\end{equation*} 
with respect to trajectories of vector states $\ket{\Psi(t)}$ in the
Hilbert space. Therefore minimizing $S_{\rm D}$ over trajectories in a
chosen variational family of states $\ket{\Psi_{\rm var}}$ gives an
approximate solution of Schr\"odinger equation. In QFT, where scalar
products are written as functional integrals, calculability strongly
reduces the allowed family of states. Choosing states represented by
Gaussian wavefunctionals, for which calculability is manifest, yields the 
HF approximation. Let us in fact consider the
wavefunctional
\begin{equation} \label{gaussstate} 
  \Psi[\varphi](t)=
  \mathcal{N} \exp\left\{ i \braket{p(t)}{\varphi-\phi(t)} - 
    \bra{\varphi - \phi(t)}\left[\tfrac14 \G^{-1}(t)+ i\s(t)\right] 
  \ket{\varphi-\phi(t)}\right\}
\end{equation}
where the variational parameters are the background field $\phi(x)$, the
background momentum $p(x)$, the real symmetric positive kernel $\G_{ij} (
\bm{x},\bm{y};t )$ and the real symmetric kernel $\s_{ij} ( \bm{x},\bm{y};t
)$ with the short-hand notation
\begin{equation}\label{notazintegr}
  \langle a (t) | M (t) | b (t) \rangle \equiv \int d^3 x\, 
  d^3 y~ a_j( \bm{x}, t )  M_{jk} ( \bm{x},\bm{y};t )~b_k ( \bm{y}, t ) 
\end{equation}
By evaluating $S_{\rm D}$ we obtain the variational action for these
parameters
\begin{equation}\label{HFaction}
  \begin{split}
    \Gamma_{\rm HF}[\phi, p, \G,\s] &= \int dt 
    \Big( \langle p(t)|\dot{\phi}(t)-p(t)\rangle - 
    \mathcal{V}[ \G(t),\phi(t) ] \\ &+
    \mathrm{Tr} \left[ \dot{\G}(t)\s(t)
    - 2 \s (t)\G(t)\s(t)- \tfrac18 \G(t)^{-1}\right]
  \end{split}
\end{equation}  
where traces are taken over all indices and space variables, and
\begin{equation*}
  \begin{split}
    \mathcal{V}\left[ \G,\phi \right] &= \frac{1}{2} \langle \phi |- \Delta +
    m_0^2| \phi  \rangle + \frac{1}{2} \mathrm{Tr}
    \left[( - \Delta + m_0^2 ) \G \right] + \int d^3 x \Big( 
    \frac{\lambda_0}{4!} ( \phi_i ( \bm{x} ) \phi_i ( \bm{x} ))^2  \\
    &+ \frac{\lambda_0}{12} \phi_i ( \bm{x} )  \phi_i ( \bm{x} ) \G_{jj} (
    \bm{x}, \bm{x}) + \frac{\lambda_0}{6} \phi_i ( \bm{x} )  \phi_j (
    \bm{x} ) \G_{ij} ( \bm{x}, \bm{x}) \\
    &+ \frac{\lambda_0}{4!} \G _{ii}( \bm{x}, \bm{x} ) \G_{jj} ( \bm{x},
    \bm{x} ) +\frac{\lambda_0}{12} \G_{ij} ( \bm{x}, \bm{x} ) \G_{ij} (
    \bm{x}, \bm{x} ) \Big)
  \end{split}
\end{equation*}
Notice that the HF Dirac action is a first order action that involves only
the first time--derivative of fields and treats coordinates and momenta as
independent variables.  The equations of motion are
obtained by variation of eq.~(\ref{HFaction}) as
\begin{equation} \label{HFeqskernels}
  \begin{split}
    \ddot{\phi}_i &= -\left\{ \left[ -\Delta+ m_0^2 +\tfrac16\lambda_0
        \,\phi_k\phi_k \right] \delta_{ij} + \tfrac12 \lambda_0\, 
      \tau_{ijkm} \G_{km} (\bm{x}, \bm{x}) \right\} \phi_j \\
    \dot{\G}_{ij} &= 2[ \G \,\s + \s \, \G]_{ij} \\
    \dot{\s}_{ij} ( \bm{x}, \bm{y}) &= [\tfrac18 \G^{-2}-2\,\s^2]_{ij}
    (\bm{x}, \bm{y}) \\ & - \tfrac{1}{2} \left[ (- \Delta + m_0^2\,)
      \delta_{ij} + \tfrac12 \lambda_0\, \tau_{ijkm} (\phi_k\phi_m +
      \G_{km}(\bm{x}, \bm{x}) \right] \delta^{3} ( \bm{x} - \bm{y}) 
  \end{split}
\end{equation}
where 
\begin{equation*}
  \tau_{ijkm} = \tfrac13( \delta_{ij} \delta_{km} +  \delta_{ik} \delta_{jm} + 
  \delta_{im} \delta_{jk} )
\end{equation*}
and matrix multiplication, over both discrete and continuous indices, is
understood. We have tacitly replaced in the last equations, and in general in
the quantum Hamiltonian, the constant parameters $\lambda$, $m^2$ with bare
(cut-off dependent) parameters $\lambda_0$, $m^2_0$. In fact the theory
should be thought as regularized with an $UV$ cut-off $\Lambda$ and then
renormalized to remove the divergent dependence on $\Lambda$. We have not
included here any field renormalization because, as is well known, this is
absent in any mean--field type approach such as the HF approximation.

To conclude this subsection we introduce an equivalent formulation of the
eqs.~(\ref{HFeqskernels}) in terms of mode functions $u_{\bm{k}\,a}$
($\bm{k}$ is the wavevector and $a$ the $O(N)$ polarization), that will be
useful later on. For simplicity let us suppose that the initial ($t=0$)
kernels are transitionally invariant (while the background field and
momentum may be point dependent). We can then write
\begin{equation*}
  \begin{split}
    \G_{ij} ( \bm{x}, \bm{y};0 ) &= \int \frac{d^3 k }{(2\pi)^3}
    \tilde{\G}_{ij} ( \bm{k} )~ e^{i \bm{k} \cdot \bm{x}} \\
    \s_{ij} ( \bm{x}, \bm{y};0 ) &= \int \frac{d^3 k }{(2 \pi)^3} 
    \tilde{\s}_{ij} ( \bm{k} )~ e^{i \bm{k} \cdot \bm{x}}
  \end{split}
\end{equation*}
Next, we introduce the $t=0$ mode functions by 
\begin{equation} \label{mfinival}
  \begin{split}
    u_{\bm{k}\,a,\,i}(\bm{x},0) &= [\tilde{\G}( \bm{k})^{1/2}]_{ai} 
    e^{i \bm{k} \cdot \bm{x}} \\
    \dot{u}_{\bm{k}\,a,\,i}(\bm{x},0) &= \left[- \tfrac{i}2 
      \tilde{\G}(\bm{k})^{-1} + 2\, \s(\bm{k})\right]_{ij} 
    u_{\bm{k}\,a,\,j} (\bm{x}, 0)
  \end{split}
\end{equation}   
and let them evolve according to the equations of motion 
\begin{equation} \label{HFeqsmf}
  \left\{ \left( \Box + m_0^2 \right) \delta_{ij} +  
    \tfrac12\lambda_0 \tau_{ijmn} 
    \left[ \phi_m(x) \phi_n(x) +  \int \frac{d^3 p }{(2\pi)^3}~ 
    u_{\bm{p}\,b,\,m}(x) \,\overline{u}_{\bm{p}\,b,\,n}(x)  
  \right]\right\} u_{\bm{k}\,a,\,j} (x) =0
\end{equation}
Then one can show that the last two equations in eqs.~(\ref{HFeqskernels}) 
are equivalent to  
\begin{equation} \label{mfkerrels}
  \begin{split}
    \G_{ij}( \bm{x}, \bm{y} ; t ) &= \int \frac{d^3 k }{(2\pi)^3}~ 
    u_{\bm{k}\,a,\,i}(\bm{x}, t)\,\overline{u}_{\bm{k}\,a,\,j}(\bm{y},t)\\
    \dot{u}_{\bm{k}\,a,\,i}(\bm{x},t)&= \int d^3 y \left\{
      - \frac{i}2 [\G^{-1}]_{ij}(\bm{x}, \bm{y} ; t ) + 2 \,\s_{ij}
      (\bm{x},\bm{y};t) \right\}u_{\bm{k}\,a,\,j}(\bm{y},t)
  \end{split}
\end{equation}
Through the mode functions we can introduce the symmetric correlation   
\begin{equation*}
  G_{ij}(x,y) = G_{ji}(y,x) =  \mathrm{Re} \int \frac{d^3 k }{(2\pi)^3}~ 
  u_{\bm{k}\,a,\,i}(x)\,\overline{u}_{\bm{k}\,a,\,j}(y) 
\end{equation*}
whose equal time value reproduce the kernel $\G$,
\begin{equation*}
  G_{ij}(x,y)\big|_{x_0=y_0=t} = \G_{ij}( \bm{x}, \bm{y} ; t )
\end{equation*}
Then we can reformulate the dynamics in terms of $\phi$ and $G$ in a
manifestly covariant way as
\begin{equation} 
\label{HFcorrfunct}
  \begin{split}
    &\left\{ \left[ \Box + m_0^2 +\tfrac16\lambda_0
        \,\phi_k(x)\phi_k(x) \right] \delta_{ij} + \tfrac12 \lambda_0\, 
      \tau_{ijkm} G_{km} (x, x) \right\} \phi_j(x)=0 \\[2mm]
 & \left\{ \left( \Box + m_0^2 \right) \delta_{ij} +  
    \tfrac12 \lambda_0 \,\tau_{ijkm} 
    \big[ \phi_k(x) \phi_m(x) +  G_{km}(x,x)  
  \big]\,\right\}  G_{ij}(x,y) =0
  \end{split}
\end{equation}


\subsection{CTP formalism and resummations}\label{CTP}
 
The general approach to non equilibrium problems in QFT was developed
by Keldish and Schwinger. It is known as closed time path (CTP)
formalism and allows to use standard path integral functional methods
(see~\cite{Calzetta:1986cq,Cornwall:1974vz,Cooper:1995zs,Chou:1984es}).
Basically it is obtained by introducing path integrals on a 
time path going from $t=0$ to $t=+\infty$ and back. Field integration
variables are then doubled and subdivided into $(+)-$components, for
the path integral forward in time, and $(-)-$components for the
backward piece.  Given an initial state defined by the functional
$\Psi[ \varphi]$ (in our case it will be a Gaussian state) one writes
down the functional integral
\begin{equation}\label{CTPgenfun}
    e^{i\,\mathcal{W}[j_+,j_-]} = \int \mathcal{D}\varphi_+ 
    \mathcal{D} \varphi_- \Psi[\varphi_+]  \overline{\Psi} [ \varphi_-]
    \,e^{\,i S[\varphi_+]-i S[\varphi_-]+ i \fbraket{j_+}{\,\varphi_+}
    -i\fbraket{j_-}{\,\varphi_-}}
\end{equation}
where $S$ is the classical action (\ref{classaction}) and we have used
the notation (\ref{notazintegr}) for the currents terms. Integration
is on trajectories from $t=0$ to $t=+ \infty$ (with the condition
$\varphi_+ = \varphi_-$ at $t= +\infty$) and $\varphi_\pm$ in the
wave functional is the $t=0$ section of $\varphi_\pm$. By construction
$\mathcal{W}[j_+,j_-]$ is the generating functional of connected Green
functions
\begin{equation*}
  \begin{split}
    G_{+\dots+-\dots-}(x_1,\dots,x_n,y_1,\dots,y_n) &\equiv
    \frac{ (-i)^{n+m}\,\delta^{n+m} \mathcal{W}}
    {\delta j_+(x_1)\dots \delta j_+(x_n)\delta j_-(y_1)
      \dots \delta j_-(y_m)} \Bigg|_{\substack{j_+ = 0 \\j_- = 0}} \\[2mm] 
    &= -i\bra{\Psi} \overline{\mathcal{T}} \{\varphi(y_1)\dots \varphi(y_m)\} 
    \mathcal{T}\{\varphi (x_1)\dots \varphi(x_n) \}\ket{\Psi}_{\rm conn}
  \end{split}
\end{equation*}
where $\mathcal{T}$ and $\overline{\mathcal{T}}$ define time ordered and inverse
ordered products, respectively, and internal indices have been omitted for ease
of notation. The effective action $\Gamma_{\rm 1PI}[\phi_+,\phi_-]$, which is
the generator of 1PI vertex functions, is the Legendre transform of
$\mathcal{W}[j_+,j_-]$ from the currents $j_{\pm}$ to the fields
$\phi_{\pm}$. The equation of motion for the background field
$\phi(x)=\bra{\Psi}\varphi(x)\ket{\Psi}$ then reads
\begin{equation} \label{eqmotCPT}
  \frac{\delta \Gamma_{\rm 1PI}}{\delta \phi_+ (x)} \Bigg|_{\phi_+=\,\phi_-=\phi}=0
\end{equation}   
Notice that, in the case of the Gaussian wavefunctional eq.~(\ref{gaussstate}),
$\Gamma_{\rm 1PI}$ depends parametrically only on the $t=0$ kernels $\G(0)$ and
$\s(0)$, while the $t=0$ background field $\phi(0)$ and $p(0)$ enter, instead,
as initial conditions for the equation eq.~(\ref{eqmotCPT}).

The perturbative diagrammatic expansion in the CTP formalism proceeds as in
vacuum QFT in terms of free propagators
\begin{equation*}
  \begin{split}
    G^0_{++} (x,y) &\equiv G^0_{F}(x,y) = -i \bra{\Psi}\mathcal{T}
    \varphi( x )\varphi( y )\ket{\Psi}_{\rm conn}  |_{\lambda=0} \\
    G^0_{--} (x,y) &\equiv G^0_{\bar F}(x,y) = -i \bra{\Psi}\overline{\mathcal{T}}
    \varphi( x )\varphi( y )\ket{\Psi}_{\rm conn}  |_{\lambda=0} \\
    G^0_{+-}(x,y) &= G^0_{-+}(y,x) = -i\bra{\Psi}\varphi( y )\varphi( x )
    \ket{\Psi}_{\rm conn}  |_{\lambda=0}
  \end{split}
\end{equation*}      
and vertices (for clarity, we write here explicitly the internal indices)
\begin{equation}
  \label{CTPvert}
  \centering
  \includegraphics[width=.5\textwidth]{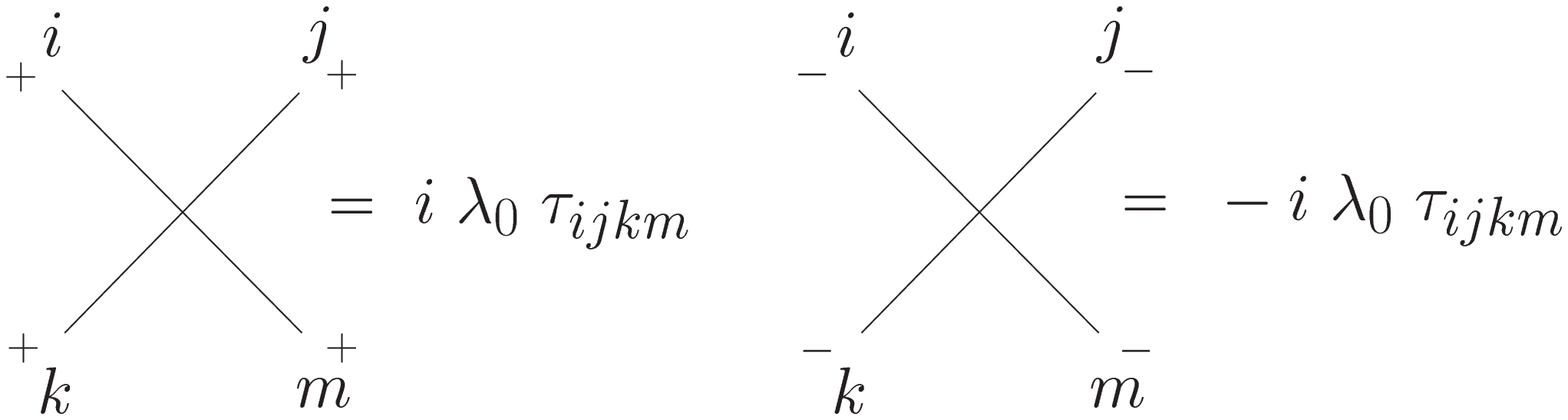}
\end{equation}
It can be shown that, in generic out--of--equilibrium contexts or even
at equilibrium with nonzero temperature, plain perturbation theory is
of very little help and many resummation methods have been developed
to go beyond it. A very successful instrument in this sense is the 2PI
effective action (see~\cite{Calzetta:1986cq,Cornwall:1974vz}). It is
defined as the double Legendre transform of the $\mathcal{W}$
generating functional with respect to the usual current one--point
$j_{\pm}$ and to the two--points current $K_{\alpha \beta}(x,y)$ coupled
through the term
\begin{equation*}
  \frac{1}{2} \int d^4 x \int d^4 y ~K_{\alpha \beta}( x, y )
  \varphi_{\alpha}(x) \varphi_{\beta} ( x )
\end{equation*}
where $\alpha, \beta = \pm $. $\Gamma_{\rm 2PI}$ is a functional of the
classical fields $\phi_{\alpha}$ and of the propagators $G_{\alpha \beta}$.
It yields two equations of motions
\begin{equation}\label{2PIEOM}
  \frac{\delta \Gamma_{\rm 2PI}}{\delta \phi_{\alpha} ( x )}\Big|_*= 0 \quad,
  \qquad \frac{\delta \Gamma_{\rm 2PI}}{\delta G_{\alpha \beta}( x,
  y)}\Big|_*= 0 
\end{equation}
Here the notation $|_*$ indicates that, by their physical meaning, the
$(\pm)$--component fields and propagators have to satisfy, on the
solutions of motion, the following relations 
\begin{equation*}
  \begin{split}
    \phi_-(x) &= \phi_-(x)=\phi(x) \\
    G_{F}(x,y)&= G_{+-}(y,x) \theta (x_0 - y_0) + G_{+-}(x,y) \theta
    (y_0 - x_0) \\
    G_{\bar F}(x,y)&= G_{+-}(x,y) \theta (x_0 - y_0) + G_{+-}(y,x) \theta
    (y_0 - x_0)
  \end{split}
\end{equation*}
Hence the system (\ref{2PIEOM}) reduces to two coupled equations for $\phi$ and
$G_{+-}$ only. Moreover, any initial Gaussian state may be absorbed in the $t=0$
term for the $j_\alpha$ and $K_{\alpha \beta}$ currents, so that, by the double
Legendre transform the initial Gaussian state disappears from the effective
action, but fixes the initial conditions on $\phi$ and $G_{+-}$. The role of
$\phi(0)$ and $p(0)=\dot\phi(0)$ is immediate, while for the kernels we have
\begin{equation*}
  \begin{split}
    G_{+-}(x,y)|_{x_0=y_0=0}&= -i\bra{\Psi}  \varphi( \bm y) \varphi(\bm x)
    \ket{\Psi} - i\,\phi(\bm y,0)\phi(\bm x,0)=  \G (\bm x,\bm y;0)\\
    \frac{\partial}{\partial y_0} G_{+-}(x,y)|_{x_0= y_0=0} &=-i \bra{\Psi}
    \pi( \bm y) \phi( \bm x )\ket{\Psi} - i\,p(\bm y,0)\phi(\bm x,0)\\ 
    &= 2 i \,[\G \s]({\bm x},{\bm
    y};0)+ \tfrac12 \delta^3 (\bm{x}- \bm{y})
 \end{split}
\end{equation*} 
Given $\Gamma_{\rm 2PI}$ at a certain perturbative loop order, if we
solve the second equation in (\ref{2PIEOM}) for a generic $\phi$ and
substitute the result $G[ \phi ]$ into the first one we obtain the
background equations of motion corresponding to a resummed
diagrammatic approximation of the 1PI effective action $\Gamma_{\rm
1PI}$.

For a scalar theory, the $\Gamma_{\rm 2PI}$ has the general form
\begin{equation*}
  \Gamma_{\rm 2PI} \left[ \phi, G \right]= S[\phi] + \frac{i}{2} \mathrm{Tr} 
  \left[ \log G \right] + \frac{i}{2} \mathrm{Tr} \left[ G_0^{-1} G \right] + 
  \Gamma_2 \left[ \phi, G \right]
\end{equation*}
Here $S$ is the complete classical action of the double time path (i.e. $S=
S_+ - S_-$). Traces are taken over all indices $i$, $\alpha$ and $x$.
$G_0^{-1}$ is the second derivative of the action in a $\phi$ background,
$\Gamma_2$ is the sum of all vacuum 2PI diagrams with $G$ propagators and
vertices defined by the classical action in a $\phi$ background. To two
loops level the diagrams contributing to the $\Gamma_2$ are the ``8''
and ``sunset'' diagrams
\begin{equation}
  \label{twoloop}
  \centering
  \includegraphics[scale=0.16]{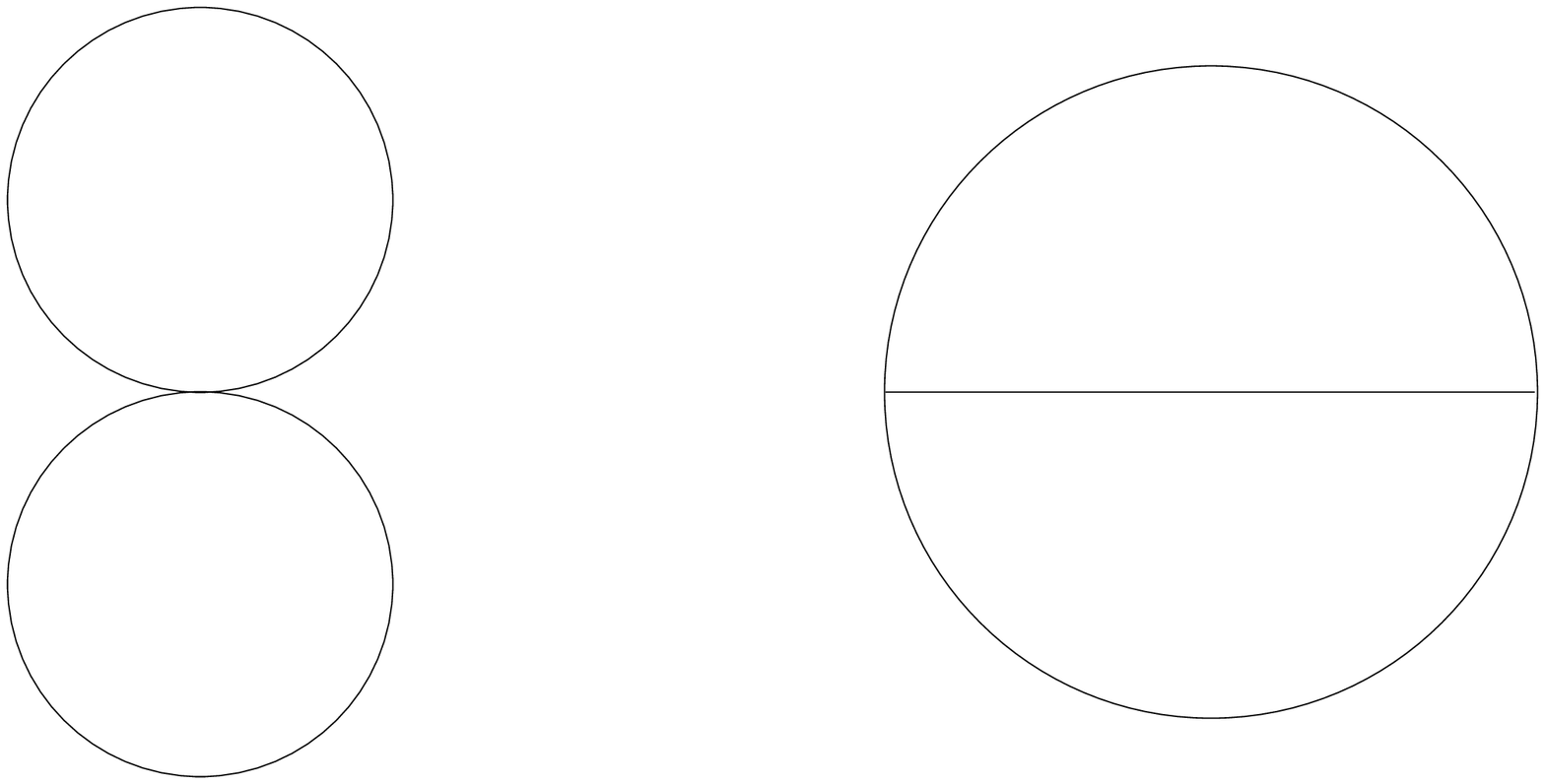}
\end{equation}
The HF approximation corresponds to consider only the first
contribution to $\Gamma_2$. 
\begin{equation*}
   \Gamma_2= \tfrac{i}{8} \lambda_0 \left[ G^2_{F} ( x,x) - G^2_{\bar F}
   ( x,x) \right] 
\end{equation*} 
In fact using this form of $\Gamma$ in the field equations (\ref{2PIEOM}),
setting $G(x,y)= \tfrac{i}{2} [ G_{-+}(x,y) + G_{-+}(y,x)]$ and observing that
the antisymmetric combination decouples, one obtains exactly the HF equations
(\ref{HFcorrfunct}).
  
Notice that ``8'' is the only 2PI diagram made of ``product'' of loops
corresponding to a mean field contribution to the mass. In the 1PI
framework this corresponds to a resummation of all vacuum 1PI diagrams
with daisy and superdaisy topologies of the form
\begin{equation}
  \label{daisygraph}
  \centering
  \includegraphics[scale=0.5]{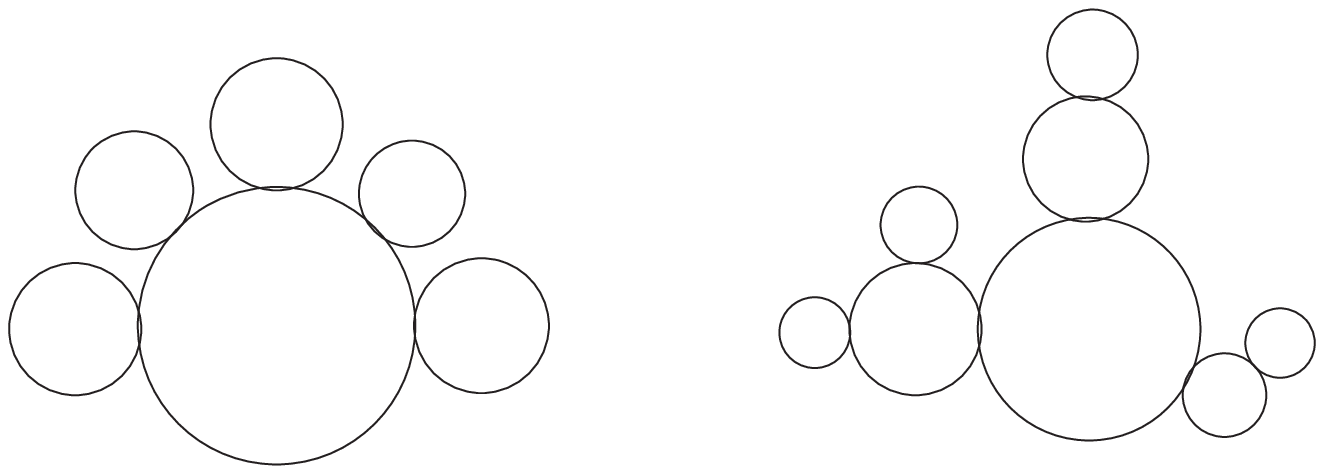}
\end{equation}

\subsection{Physical representation and reparametrization of HF effective
action}\label{PRCTP} 

In this subsection we derive useful explicit relations connecting the
diagrammatic definition of $\Gamma_{\rm HF}$ to the variational
approach of subsection~\ref{varHF}. To this purpose it is convenient
to introduce a different representation of the CTP formalism, known as
the physical representation (see~\cite{Chou:1984es}).  We introduce
the field redefinitions (omitting again the internal indices to
simplify notation)
\begin{equation*}
  \phi_{\Delta} = \phi_+ - \phi_- \quad,\qquad
  \phi_{c} = \frac{1}{2} ( \phi_{+} + \phi_{-} )
\end{equation*}
and write the 1PI effective action as a functional of these new fields,
$\Gamma_{\rm 1PI}= \Gamma[\phi_\Delta,\phi_c]$. By calculating vertex
functions, one then finds
\begin{equation}\label{PRprop1}
  \frac{\delta^{n} \Gamma}{\delta \phi_c ( x_1 ) \dots 
    \delta \phi_c ( x_n )} \Bigg|_{\phi_\Delta = 0}= 0
\end{equation}
and 
\begin{equation}\label{PRprop2}
  \frac{\delta^{n+m} \Gamma}{\delta \phi_c ( x_1 ) \dots \delta 
    \phi_c ( x_n )\delta \phi_\Delta ( y_1 ) \dots \delta \phi_\Delta 
    ( y_m )} \Bigg|_{\phi_\Delta = 0}= 0
\end{equation}
if the time component of anyone of the $x$'s is larger than the time
component of anyone of the $y$'s. Let us also remark that, by the
definition of CTP generating functional, all time coordinates in the
vertex functions are supposed to be positive so we can set, as well,
these functions to be zero for any negative time. Using
eq.~(\ref{PRprop1}) together with eq.~(\ref{eqmotCPT}) one can write
the equations of motions in the form
\begin{equation} \label{PREOM}
  \frac{\delta \Gamma}{\delta \phi_{\Delta}} \Bigg|_{\phi_\Delta=0,~\phi_c=\phi}=0
\end{equation}
Notice that $2n$--legs vertex functions with one $\phi_\Delta$
leg and $2n-1$ $\phi_c$ legs are the only ones contributing
to these equations of motion.  Then eq.~(\ref{PRprop2}) guarantees that
all the terms nonlocal in time in eq.~(\ref{PREOM}) do satisfy
causality.

Perturbative calculations by diagrammatic expansion in the physical
representation are based on the free propagators
\begin{equation*}
  \begin{split}
    G^0_{c \Delta} (x,y) &\equiv G^0_{A}(x,y) = -i \theta ( y_0 - x_0 )
    \bra{\Psi}[\phi( x ),\phi( y )]\ket{\Psi}  |_{\lambda=0} \\
    G^0_{\Delta c} (x,y) &\equiv G^0_{R}(x,y) = -i \theta ( x_0 - y_0 )
    \bra{\Psi}[\phi( x ),\phi( y )]\ket{\Psi}  |_{\lambda=0} \\
    G^0_{\Delta \Delta} (x,y) &= -2i \bra{\Psi}
    \{\phi( y ),\phi( x )\}\ket{\Psi}_{\rm conn}  |_{\lambda=0} \;,\quad
    G^0_{c c} (x,y) = 0
  \end{split}
\end{equation*} 
The free retarded and advanced Green functions $G_A^0$ and $G_R^0$ do
not depend on the initial state and are translational invariant.  The
correlation function $G=\tfrac{i}{4}G_{\Delta \Delta}^0$,
instead, does depend on $\ket{\Psi}$. The vertices are
\begin{equation}
  \label{PRvertices}
  \centering
  \includegraphics[width=0.5\textwidth]{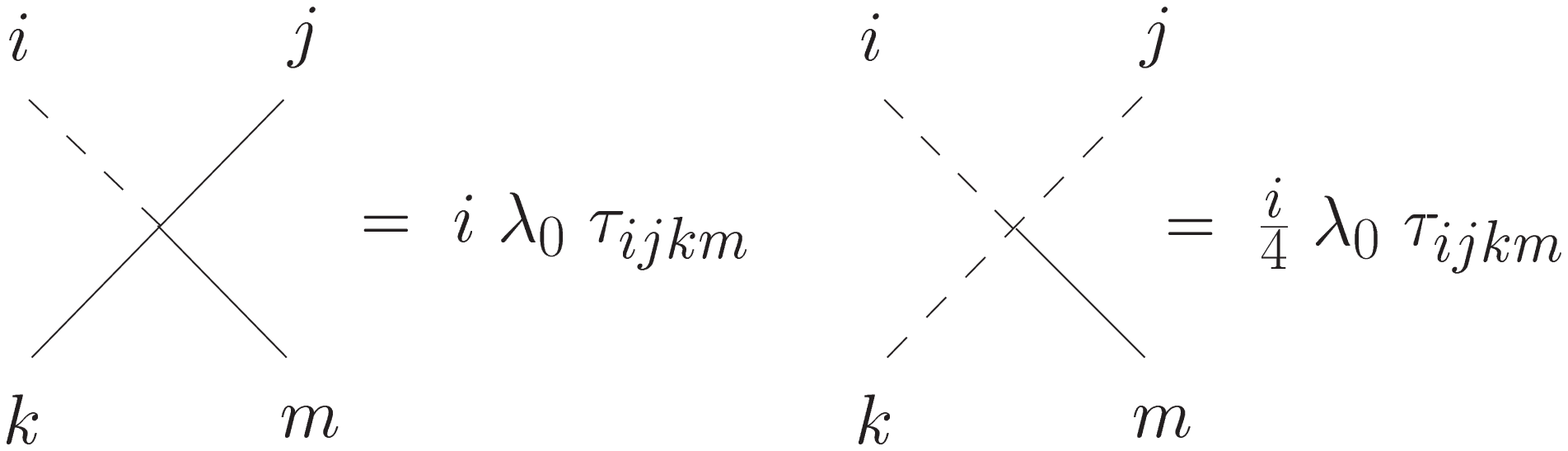}
\end{equation}
where solid lines represent $\phi_c$ legs while dotted lines represent
$\phi_\Delta$ legs.

As stated above, the HF approximation consists in the
resummation of diagrams with daisy and superdaisy topologies. Therefore the
functional differentiations which produce the vertex functions act only on
propagators; hence the diagrams contributing to a $2n$-legs vertex function
(both in $\pm$ and physical representations) must have the form
\begin{equation*}
  \centering
  \includegraphics[scale=0.4]{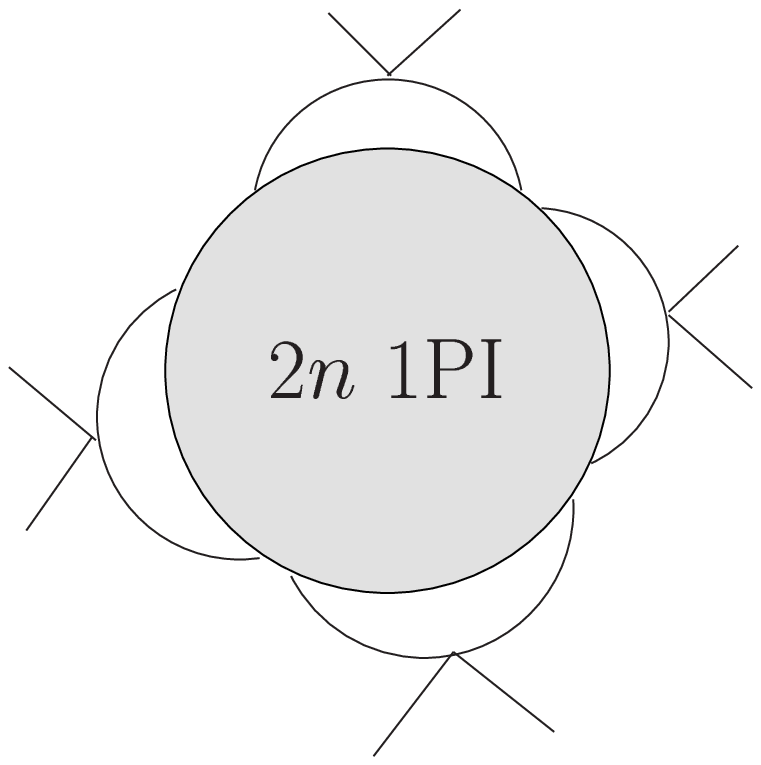}
\end{equation*}
that is, the $2n$ legs are pairwise connected to to $n$ vertices. 
This means that the HF effective action can be written in the form
\begin{equation*}
  \Gamma_{\rm HF}[\phi_\Delta, \phi_c]= -\bra{\phi_\Delta}\Box\ket{\phi_c} - 
  \F[ \xi, \chi, \eta]
\end{equation*}
where $\F$ is a functional of the following composite matrix fields 
\begin{equation*}
  \begin{split}
    \xi_{ij} ( x ) &= \phi_{c,i}( x )\phi_{c,j}( x )\\
    \chi_{ij} ( x ) &= \phi_{c,i}( x )\phi_{\Delta,j}( x )\\
    \eta_{ij} ( x ) &= \phi_{\Delta,i}( x )\phi_{\Delta,j}( x )
  \end{split}
\end{equation*}
We now introduce, for ease of notation, the new object
\begin{equation*}
  \F'[\xi]_{ij} ( x ) = \frac{1}{2}
  \frac{\delta \F}{\delta \chi_{ij}(x)} \Bigg|_{\chi=\eta=0}
\end{equation*}
which is a functional of
$\xi_{ij}(x)=\phi_{c,i}(x)\phi_{c,j}(x)=\phi_i(x)\phi_j(x)$ only. Then
the equation of motion in the HF approximation takes the general form
\begin{equation}\label{HFEOMgeneralform}
  \big( \Box \delta_{ij} + 2 \,\F'_{ij}\big) \phi_j=0
\end{equation}
By comparison with eq.~(\ref{HFeqskernels}) and eq.~(\ref{mfkerrels}) we
obtain the following expression for $\F'$
\begin{equation}\label{Gammaformula}
  \F'[\xi]_{ij}( x )= \frac{m_0^2}{2} \delta_{ij} + 
  \frac{\lambda_0}{12}\delta_{ij} \xi_{kk} ( x ) + \frac{\lambda_0}{12}
  t_{ij}^{km}  \int \frac{d^3 p }{(2 \pi)^3}\, 
  u_{\bm{p}\,a,\,i}(x) \, \overline{u}_{\bm{p}\,a,\,j}(x)
\end{equation} 
The dependence of the mode functions $u_{\bm{p}\,a}$ on $\xi$ is fixed
by solving eq.~(\ref{HFeqsmf}) for a generic $\phi$ field. Notice that
the identification of eq.~(\ref{Gammaformula}) is somewhat arbitrary
since we can add to $\F'$ a generic functional term
$A[\xi]_{icky}(x) \xi_{km}(x)$ with $A_{icky}=-A_{skim}$ without
changing the contribution of $\F'$ to the effective action and to the
equations of motion~(\ref{HFEOMgeneralform}).

The role of the functional $\F'$ in the generation of
vertex functions can be easily established. Let us consider the
$2n$-points vertex function with one $\phi_\Delta$ field leg and
$m(=2n-1)$ $\phi_c$ field legs (as we said those we are interested in
for they contribute to the equations of motion), a simple calculation
leads to
\begin{equation}\label{2Eptfunct}
 \begin{split} \Gamma^{(2n)}_{i|j_1 \dots j_m} (x|y_1,\dots,y_m )
   &\equiv   \frac{\delta^{2n} \Gamma}{\delta
     \phi_{\Delta~i}(x) \delta \phi_{c~j_1}(y_1) \dots \delta
     \phi_{c~j_m}(y_m)}\Bigg|_{\substack{\phi_\Delta=0\\\phi_c=0}} \\[4mm]
   &= \Box~ \delta^4 ( x - y_{j_1})\, \delta_{1,n}+ \frac{2}{(n-1)!}\sum_{p \in
   \Sigma_m} \delta^4 (x - py_1)\\ & \times\left[\prod_{k=1}^{n-1}
   \delta^4(py_{2k} - py_{2k+1})\right]\frac{\delta^{n-1}
   \F'[\xi]_{i\, pj_1}(x)}{\delta p\xi_{j_2 j_3} ( y_2)
   \dots \delta p\xi_{j_{m-1} j_m} ( y_m)}\Bigg|_{\xi=0}
   \end{split}
\end{equation}
Where $\Sigma_m$ is the set of all permutations on $\{ 1 \dots m \}$
and we used the notation $py_i \equiv y_{p(i)}$, $pj_i \equiv
j_{p(i)}$  and $p \xi_{j_i j_k}(y_s) \equiv \xi_{pj_i \, pj_k}(py_s)$.

In what follows we are going to study the problem of renormalizability (and
RG--invariance) of the effective field equations in the HF approximation. By the
results of this section we need to consider only the vertex functions
contributing to the equations of motion and hence only the functional
$\F'$.

\section{Renormalizability and RG invariance}\label{Renormalization}

We are concerned here about the possibility of constructing a set of HF
equations in which all the divergent dependence from the cut-off has been
removed by a suitable renormalization procedure and, at the same time, no
dependence on renormalization scale has been introduced. In brief we are going
to study the problem of renormalizability and RG invariance of HF equations.
This will lead us to some results that go beyond the strict mean field
approximation and include resummations coming from 2PI effective action loop
expansion.

In this section we will deal with the general case of an inhomogeneous (in
space) dynamical problem within the $O(N)$ symmetric scalar field theory with
unbroken symmetry. However the main aspects of the problem can be more easily
pointed out by considering a simpler case. In the first subsection, therefore,
we restrict to the static problem, that is the calculation of the effective
potential, for the $N=1$ theory.

\subsection{A simple instructive example: effective potential for
$N$=1} \label{effpot}  

We want to study the renormalization problem of the static and translationally
invariant version of eqs.~(\ref{HFeqskernels}) when $N=1$. This situation
follows, in the variational approach of sec.~\ref{varHF}, by considering a field
$\phi$ constant throughout space--time, with vanishing momentum $p$, and a
constant translationally invariant $\G$ kernel
\begin{equation*}
  \G ( \bm{x}-\bm{y} ) = \int \frac{d^3 k }{(2 \pi)^3} \,
  \tilde \G ( \bm{k} )\, e^{-i \bm{k}( \bm{x} - \bm{y} )}
\end{equation*}
with $\s =0$. By a simple substitution in eq.~(\ref{HFeqskernels}) we
obtain
\begin{equation}\label{staticeqs}
  \begin{split}
    \frac{1}{4 \tilde\G^2(\bm k)} &= \bm{k}^2 + m_0^2 + 
    \frac{\lambda_0}{2} \phi^2 + 
    \frac{\lambda_0}{2}\int_{|\bm p|<\Lambda} \frac{d^3 p }{(2 \pi)^3}\, 
    \tilde \G ( \bm{p} )\\ 0 &= \left[ m_0^2 + 
      \frac{\lambda_0}{6} \phi^2 + \frac{\lambda_0}{2}
    \int_{|\bm p|<\Lambda} \frac{d^3 p }{(2 \pi)^3}
    \,\tilde \G ( \bm{p} )\right] \phi
  \end{split}
\end{equation}
where a sharp cut-off $\Lambda$ has been introduced as regularization.

Equivalently we can consider the equations of motion as defined by the 1PI
effective action and reduce to the static case by evaluating 
$\Gamma_{\rm 1PI}$ on a space--time constant $\phi$. So letting
\begin{equation*}
  \hat{V}' (\xi)= \F'[\xi]\big|_{\xi=\mathrm{constant}}
\end{equation*}
we can define the effective potential by $V_{\rm eff}(\phi)=\hat{V}
(\phi^2)$, with $\hat{V}(\xi)$ the primitive of $\hat{V}'(\xi)$
vanishing at $\xi=0$. From eq.~(\ref{HFEOMgeneralform}) the equation for
the background field $\phi$ is
\begin{equation*}
  V_{\rm eff}'(\phi) = 2\,\hat{V}' ( \phi^2 ) \,\phi =0
\end{equation*}
Comparison with the second equation in (\ref{staticeqs}) leads immediately to
\begin{equation}\label{effpoteq1}
  \hat{V}' (\xi)  = \frac{m_0^2}{2} + \frac{\lambda_0}{12} \xi + 
  \frac{\lambda_0}{4}\int_{|\bm p|<\Lambda}
  \frac{d^3 p }{(2 \pi)^3}\,\tilde \G (\bm{p})
\end{equation}
where the implicit dependence of $\tilde \G$ on $\xi=\phi^2$ is obtained by
solving the first eq.~(\ref{staticeqs}) with a generic $\phi$. We recall
also that $V_{\rm eff}(\phi)$ is the generator of vertex functions with all
incoming momenta set to zero (for constant homogeneous fields the
distinction between CTP and vacuum standard formalism is immaterial).

As usual, one introduces the ansatz 
\begin{equation*}
  \tilde\G ( \bm{k} )=\frac{ 1}{ 2 \sqrt{ \bm{k}^2 + M^2 }}
\end{equation*}
that allows to cast the first of eqs.~(\ref{staticeqs}) in the form of a 
mass gap equation
\begin{equation}\label{Massgapeq}
  M^2 (\xi) = m^2_0 + \frac{\lambda_0}{2}\,\xi +  \frac{\lambda_0}{2}
  \int_{p^2<\Lambda} \frac{d^4 p}{(2 \pi)^4} \,\frac{1}{p^2 + M^2 ( \xi )}
\end{equation} 
where the integration has been written in four Euclidean dimensions ($p^2=
p_0^2+\bm{p}^2$). By comparing eq~(\ref{Massgapeq}) with
eq.~(\ref{effpoteq1}) we can rewrite the latter as
\begin{equation}\label{potandM}
  \hat{V}' ( \xi )=\frac{M^2 ( \xi )}{2} - \frac{\lambda_0}{6}\, \xi
\end{equation}
so that, by combining eq.~(\ref{Massgapeq}) and eq.~(\ref{potandM}), we have
\begin{equation}\label{eqpot}
  \hat{V}' ( \xi )=\frac{m^2_0}{2} + \frac{\lambda_0}{12} \xi + 
  \frac{\lambda_0}{4}\int_{p^2<\Lambda^2}\frac{d^4 p }{(2 \pi)^4} \frac{1}{p^2 + 
    2\hat{V}' ( \xi )+\frac13 \lambda_0\, \xi }
\end{equation}
This is a self--consistent definition of the effective potential and is the
central equation that we are going to use in this subsection.

Let us apply the standard renormalization procedure. We introduce two physical
parameters, the renormalized squared mass $m^2$ and the coupling constant
$\lambda$, which are identified, respectively, as the second and fourth
derivatives of the effective potential at zero field, so that  
\begin{equation}\label{reneq}
    \frac{d^2 V_{\rm eff}}{d\phi^2}\Big|_{\phi=0} =2\,\hat{V}'(0)=m^2 
    \quad,\qquad
    \frac{d^4 V_{\rm eff}}{d \phi^4}\Big|_{\phi=0} =12\,\hat{V}''(0)=\lambda 
\end{equation}
These are the two renormalization conditions that, once inverted, determine the
dependence of the bare parameters $\lambda_0 ( \lambda , m , \Lambda ) $ and
$m_0^2 ( \lambda , m , \Lambda )$ on the physical ones and the cut-off. The
assumed existence and positivity of $2\hat{V}'(0)$ provides the definition
of unbroken symmetry, which is our choice here.  

Using eq.~(\ref{eqpot}), we can calculate the explicit HF form of
eq.~(\ref{reneq})
\begin{equation}\label{barepar}
  m^2 = m^2_0 + \frac{\lambda_0}{2} I_E^{(1)}(m^2, \Lambda ) \quad,\qquad
  \lambda = \frac{3 \lambda_0}{1 + \tfrac12{\lambda_0} 
    I_E^{(2)}(m^2, \Lambda )}- 2 \lambda_0 
\end{equation}
where, to shorten notation, we have introduced  
\begin{equation}\label{integral}
  I^{(n)}_E ( m^2, \Lambda ) = \int_{p^2<\Lambda^2} \frac{d^4 p }{(2 \pi)^4}
  \frac{1}{(p^2 + m^2 )^n}
\end{equation}
and the suffix $E$ refers to the Euclidean form of the integral. As
$\Lambda\to\infty$ one sees that $I^{(1)}_E$ is quadratically divergent in
$\Lambda$, while $I^{(2}_E$ is logarithmically divergent and all other 
$I^{(n)}_E$ are finite.

We now point out the problems that occur with this standard procedure.

First of all we can see that there is a pathological behaviour of the
effective quartic coupling $\lambda$ as a function of the bare parameter
$\lambda_0$ at fixed cut-off $\Lambda$. In fact $\lambda$ has the correct
$1-$loop $\lambda_0^2$ term dictated by perturbation theory, but certainly
fails at higher orders, since is exhibits an unphysical behaviour, growing to a
maximum value at $\lambda_0=\lambda_0^{max}$, then decreasing to zero and
to even more unphysical negative values. This implies the breakdown of
the HF approximation for values of $\lambda_0$ greater than
$\lambda_0^{max}$ in a theory at fixed cut-off. Another way of looking at
this problem is to invert eqs.~($\ref{barepar}$) to obtain $m^2_0$ and
$\lambda_0$ as functions of $\Lambda$ parametrized by $m^2$ and $\lambda$.
A full trajectory in $\Lambda$ of $m^2_0$ and $\lambda_0$ corresponds to a
single renormalized theory describing the dynamics at momentum scales much
smaller than $\Lambda$. Clearly we must restrict our attention to the
monotonically increasing branch through the origin at $\lambda_0=\lambda=0$.
But these trajectories, for any positive value of $\lambda$,
exhibit a ``Landau obstruction'' at the value of $\Lambda$ which corresponds to
$\lambda_0^{max}$. This obstruction is even more troublesome than the Landau pole
present in the standard $1-$Loop--Renormalization--Group improved relation,
which by a suitable choice of finite parts can be written as  
\begin{equation}\label{RGI1Lcoupl}
  \lambda_0(\lambda, \Lambda/m)\big|_{\rm 1LRG} = \frac{\lambda}
  {1 - \tfrac32 \lambda\, I_E^{(2)}(m^2, \Lambda )} 
\end{equation}
in terms of the whole $I_E^{(2)}(m^2,\Lambda)$ rather than just its leading
divergence $\frac1{8\pi^2}\log(\Lambda/m)$. In fact, the obstruction spoils even
the one--to--one correspondence between bare and renormalized parameters (at
fixed cut--off) which must hold true in general and holds true also in the 1LRG
improved relation (\ref{RGI1Lcoupl}).

Secondly, and even more seriously, one may verify that the above
renormalization procedure does not remove all the logarithmic UV--cutoff
dependence from the potential. That is, $V_{\rm eff}(\phi)$ does not
parametrically depend solely on $m^2$ and $\lambda$, but also on $\lambda_0$
and therefore on $\log\Lambda$. In a QFT with a Landau pole or
obstruction, where the UV cutoff cannot be completely removed, we should
accept at most an inverse power dependence of physical quantities on
$\Lambda$.

To show this, consider the finite part of $I_E^{(1)}$, that is
\begin{equation*}
  J ( y, \Lambda ) = I_E^{(1)}( y, \Lambda ) - I_E^{(1)}( m^2, \Lambda ) 
  -  I_E^{(2)}( m^2, \Lambda )(m^2-y)
\end{equation*}
for any real $y$. By the use of $J$ eq.~(\ref{eqpot}) may be cast in the
form
\begin{equation*}
  \hat{V}'(\xi)= \tfrac12 m^2 + \tfrac1{12}\lambda\,\xi + 
  \tfrac1{12}(\lambda + 2\,\lambda_0) \,J \big( 2\,\hat{V}'(\xi)+
    \tfrac13\lambda_0 \xi - m^2 ,\Lambda \big)
\end{equation*}
This can be written more compactly 
\begin{equation}\label{eqpotQR2}
  \Delta \hat{V}' ( \xi )= \tfrac14{g} \, J\big(2\,\Delta \hat{V}'(\xi) + 
  \tfrac12 g\, \xi \big)
\end{equation} 
with the definitions 
\begin{equation}\label{gdef}
    g = \tfrac13(\lambda +2\,\lambda_0)\quad,\qquad  
    \Delta \hat{V}' (\xi) = \hat{V}'(\xi) - \hat{V}' ( 0 ) - 
    \hat{V}'' ( 0 )\,\xi
\end{equation}
It is quite clear now that the higher order derivatives (or vertex
functions with vanishing incoming momenta) of $V_{\rm eff}$, which are all
contained in $\Delta \hat{V}'$, depend explicitly on logarithms of the
cut-off through the effective coupling $g$.

This problem was dealt with in \cite{Destri:1999he,Pi:1987df} by a direct
renormalization of HF gap equation, by setting
\begin{equation*}
  \begin{split}
    M^2 &= m^2 + \tfrac12\lambda_0 \xi + 
    \tfrac12\lambda_0 \,I_E^{(1)} ( M^2, \Lambda )\\[1.5mm]
    &=  m^2 + \tfrac12\lambda\, \xi + \tfrac12\lambda\,J ( M^2, \Lambda ) 
  \end{split}
\end{equation*}
so that the bare--to--renormalized relations are found to be
\begin{equation}\label{Manbarepar}
  \lambda_0 = \frac{\lambda}{1-\frac12 \lambda \,
    I_E^{(2)} ( m^2, \Lambda )} \quad,\qquad
  m^2_0 = m^2 -\tfrac12\lambda_0\, I_E^{(1)} (m^2, \Lambda )  
\end{equation}
This procedure makes $M^2(\xi)$ and all higher derivatives of $\hat
V(\xi)$ cutoff independent by definition, but fails in two respects. First
of all the $1-$loop beta function read out from eq.~(\ref{Manbarepar}) is
three time smaller than the correct one [see eq.~(\ref{RGI1Lcoupl})], which
implies a mismatch already to first order in perturbation theory. Moreover
the parametrizations (\ref{Manbarepar}) do not eliminate completely the
cutoff logarithms, which remain in the quartic term of $V_{\rm eff}(\phi)$
since the background field equation reads
\begin{equation}\label{Manfrephieq}
  V_{\rm eff}'(\phi) = \big[\tfrac12 M^2(\phi^2)-\tfrac16\lambda_0
  \phi^2 \big] \phi = 0
\end{equation}
One could adjust the mismatch with the beta function by redefining the 
bare quantities appearing in the HF gap equation. If we make the ad
hoc replacements
\begin{equation}\label{Manbareparid}
  \begin{split}
    \lambda_0 &\longrightarrow \tilde{\lambda}_0 = 
    \frac{\lambda_0}{1+\lambda_0 \,I_E^{(2)} ( m^2, \Lambda )}\\
    m^2_0 &\longrightarrow {\tilde m}^2_0 = m^2 -
    \tfrac12\tilde{\lambda}_0\, I_E^{(1)} (m^2, \Lambda ) 
  \end{split}
\end{equation}
then the correct 1LRG relation (\ref{RGI1Lcoupl}) is recovered.
As for the problem with the quartic term in $V_{\rm eff}(\phi)$, 
a reasonable argument was put forward in \cite{Destri:1999he} to justify the 
substitution $\tilde{\lambda}_0 \to \lambda$ in eq.~(\ref{Manfrephieq}).

We will see that this substitution, as well as the redefinition in
eq.~(\ref{Manbareparid}), can be fully justified in the spirit of the 1LRG
improvement of the HF approximation by an explicit diagrammatic analysis.

The problems just outlined, have, in fact, a simple interpretation if
we analyze the diagrams exactly resummed by the HF approximation. First, if
we consider the renormalized mass, given by the 2 legs vertex
function with no momenta $\hat{V}' ( 0 )$, we see, by the second
equation in~(\ref{barepar}) that it is dressed by daisy and superdaisy
tadpole diagrams, as
\begin{equation}
  \label{massren}
  \centering
  \includegraphics[width=0.7\textwidth]{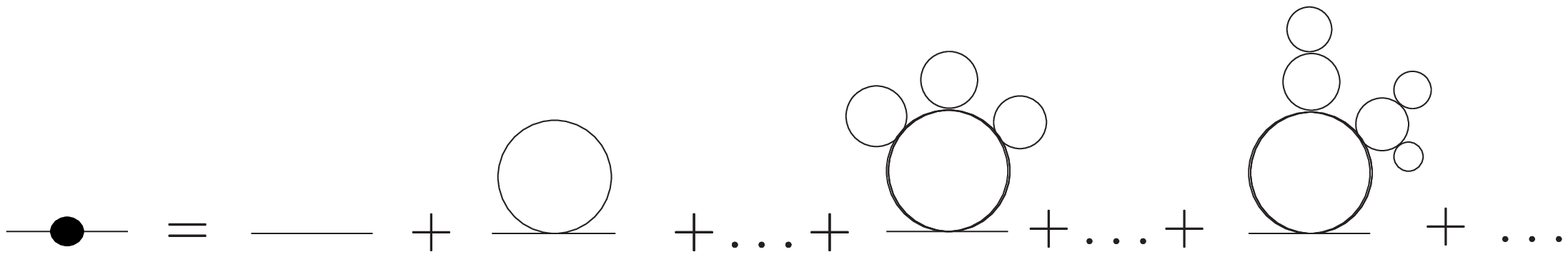}
\end{equation}  
From now on we are going to work with this dressed mass and we will 
draw a simple solid line the corresponding dressed propagator.  
Let us consider the quartic coupling, the first equation in
(\ref{barepar}) tells us it is obtained by ``chain'' diagrams (with
dressed propagators).
\begin{equation}
  \label{Coupling}
  \centering
  \includegraphics[width=0.65\textwidth]{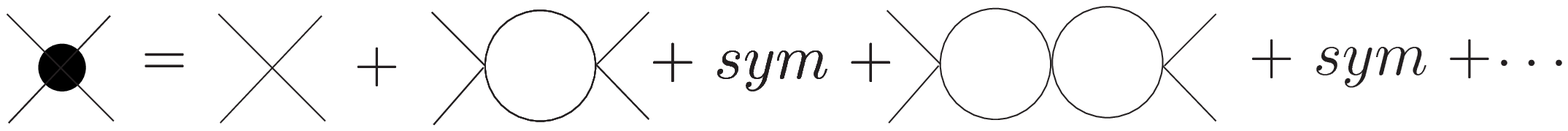}
\end{equation}
This is not the full diagrammatic expansion that corresponds to the 1LRG
relation between $\lambda$ and $\lambda_0$ in eq.~(\ref{RGI1Lcoupl}). The latter
in fact includes all local contributions with Leading Logs (LL) of the cutoff
from all diagrams at every given loop order. So eq.~(\ref{Coupling}) misses the
LL local parts of graphs like
\begin{equation}
  \label{RGIcoupling}
  \centering
  \includegraphics[width=0.75\textwidth]{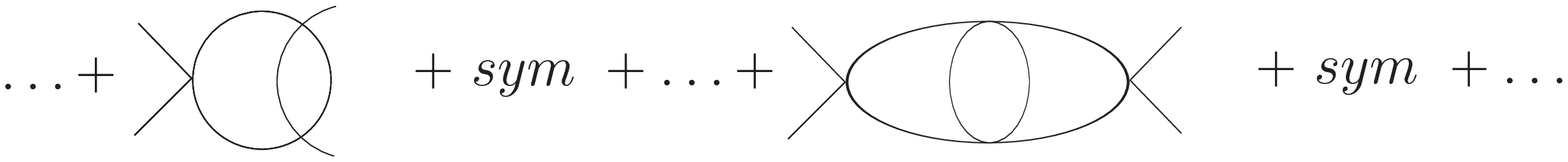}
\end{equation}
This is the cause of the problems outlined above with the form
eq.~(\ref{barepar}) of the renormalized coupling $\lambda$.

Moreover, we can examine in the same spirit the diagrams contributing to higher
order vertex functions (higher order derivatives of $\hat{V}$ in zero field).
From eq.~(\ref{eqpot}) we see that they are convergent loop diagrams
(loop with more than 2 propagators) with dressed propagators
\begin{equation}
  \label{higerorder}
  \centering
  \includegraphics[scale=0.4]{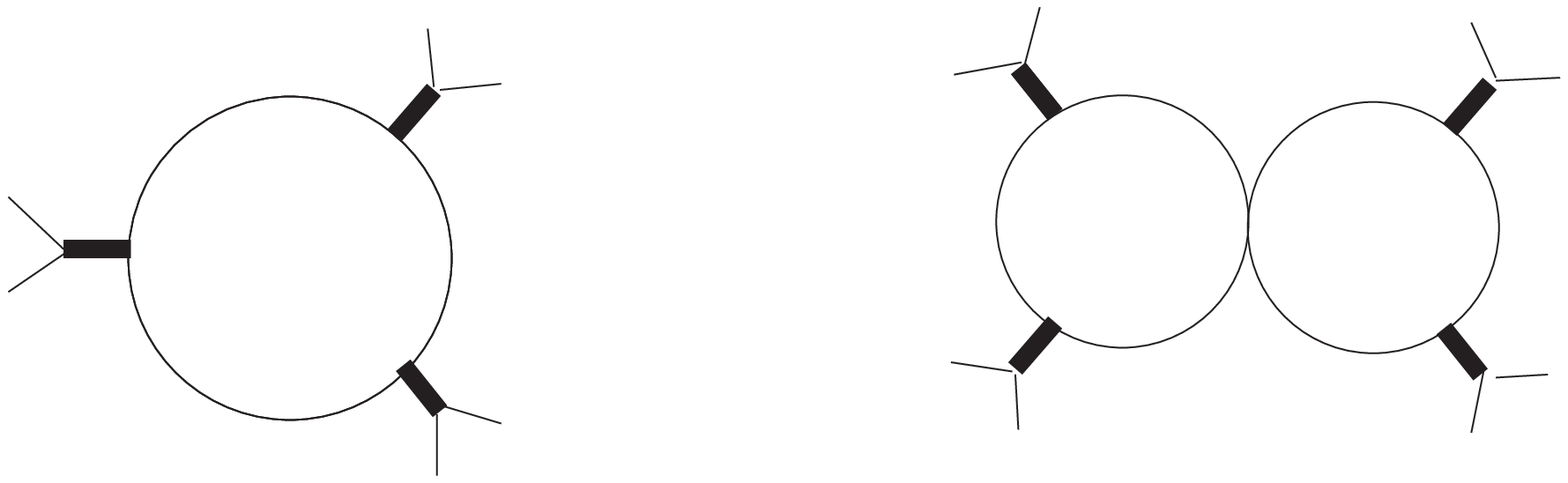}
\end{equation}
and vertices with the effective quartic coupling $g = ( 2 \lambda_0 +
\lambda )/3$
\begin{equation}
  \label{gcoupling}
  \centering
  \includegraphics[width=0.5\textwidth]{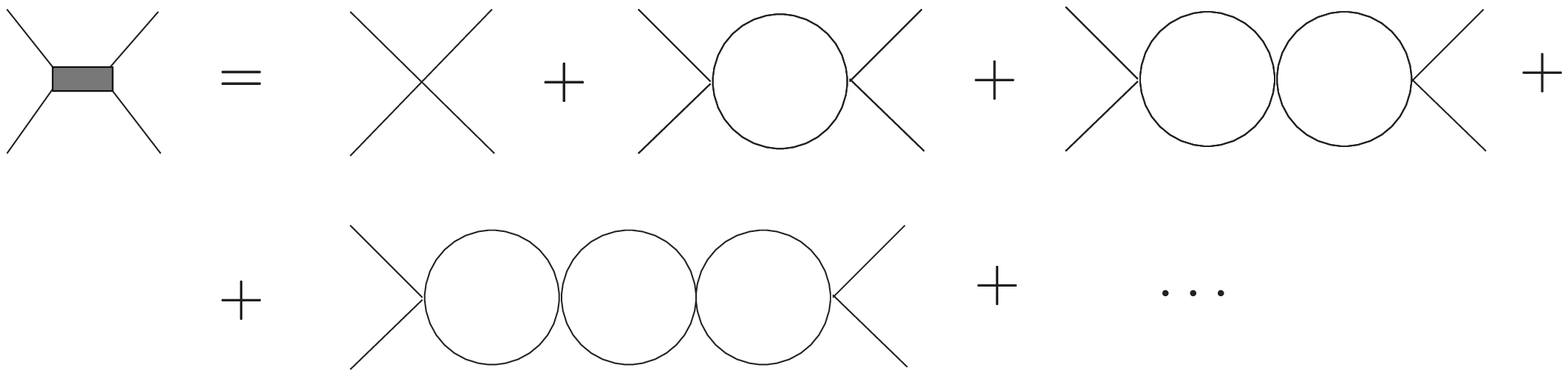}
\end{equation}
The reason of the presence of $g$ and not of the renormalized coupling
$\lambda $ in higher order vertices is that the local LL
contributions of diagrams like 
\begin{equation}
  \label{missdiagr1}
  \centering
  \includegraphics[scale=0.4]{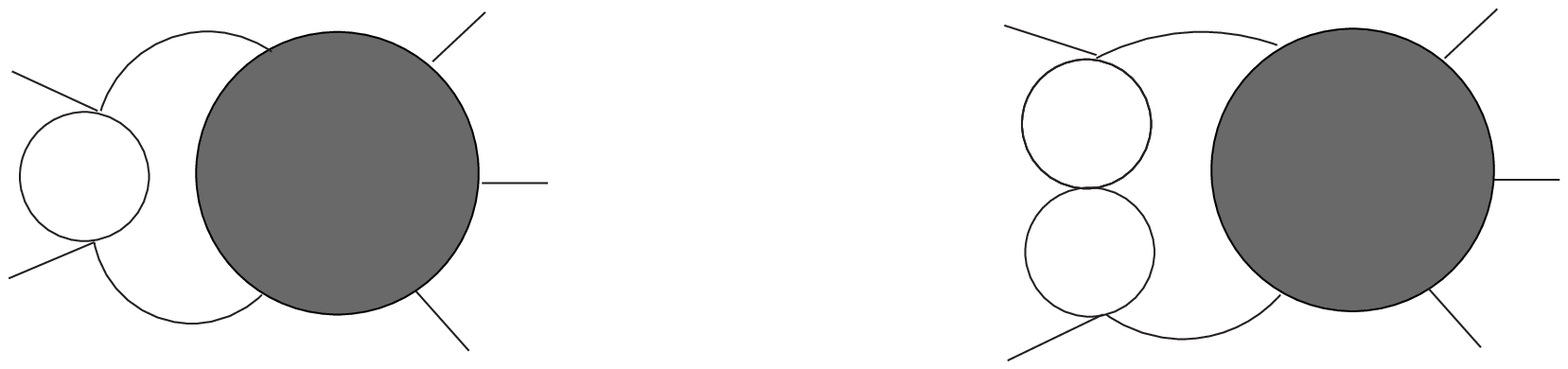}
\end{equation}
are not included in the resummation defined by the standard HF
approximation as described by eq.~(\ref{eqpotQR2}). Including such LL
contributions would make the HF approximation renormalizable in terms of the 
reparametrizations of eq.~(\ref{barepar}), that is with the pathological
relation between $\lambda$ and $\lambda_0$. This is because the LL contributions 
of diagrams like
\begin{equation}
  \label{missdiagr2}
  \centering
  \includegraphics[scale=0.4]{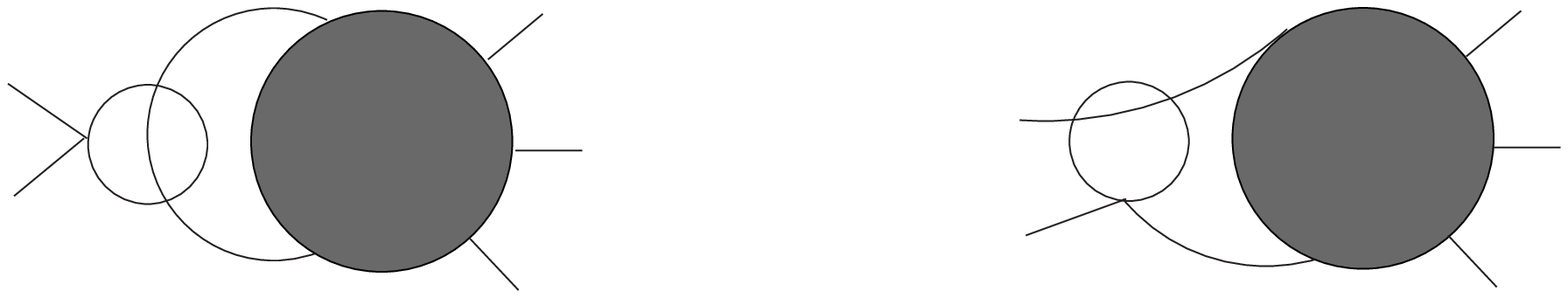}
\end{equation}
are still missing. Their proper inclusion would allow to obtain everywhere the
renormalized 1LRG improved coupling constant.

By these observations we are led to a natural way to modify the HF approximation
in order to obtain a renormalized effective potential and the
form~(\ref{RGI1Lcoupl}) for the coupling constant, that will ensure RG
invariance in the general case, as we will see. The prescription is to include
all missing contributions of the type~(\ref{RGIcoupling}) to
$12\,\hat{V}''(0)$ in order to obtain the 1LRG relation between $\lambda$ and
$\lambda_0$ as in eq.~(\ref{RGI1Lcoupl}), and to include also the contributions
of graphs of the form ~(\ref{missdiagr1}) and (\ref{missdiagr2}) in order to
obtain $\lambda$ instead of $g$ in higher order vertex functions.

In conclusion we define a modified version of HF approximation by setting
\begin{equation}\label{Veffmod}
  \begin{split}
    2 \hat{V}' ( 0 ) &= m^2 = m_0^2 + \tfrac12 \lambda_0 I_E^{(1)} ( m^2, 
    \Lambda ) \\[2mm]
    12 \hat{V}'' ( 0 ) &=  \lambda = \frac{\lambda_0}{1+\tfrac32 \lambda_0
      I_E^{(2)} ( m^2, \Lambda )}  \\[2mm]
    \Delta \hat{V}' ( \xi )&= \tfrac14 \lambda ~ J \big( 2\Delta 
      \hat{V}'( \xi ) + \tfrac12 \lambda \,\xi \big) 
\end{split}
\end{equation}
This defines to the renormalized potential $\hat{V} ( 0 )$ and leads
to the following equations
\begin{equation*}
  \begin{split} M^2 &= \tfrac12 \lambda\xi + m^2 + \tfrac12 \lambda
    J ( M^2) \\[2mm] 0 &= \left( \tfrac12 M^2 - \tfrac16
    \lambda \,\xi \right) \phi \end{split}
\end{equation*}
these are finite and exactly in the form found in
\cite{Destri:1999he}. This gives a simple diagrammatic explanation of
the substitution $\lambda_0 \to \lambda$ made in \cite{Destri:1999he}
and of the statement~(\ref{Manbareparid}).

\subsection{The general case}\label{general}
We are now in a position to deal with the general renormalization problem
of the time-dependent and inhomogeneous HF approximation for
$O(N)-$symmetric theory, as described by
eqs.~(\ref{HFeqskernels}),~(\ref{HFeqsmf}). 

Our first step is to derive a self-consistent functional equation for the
effective action analogous to the the one for the effective
potential in eq.~(\ref{eqpot}). Let us introduce the squared mass parameter
$m^2$. We choose $m^2$ to be the equilibrium mass (i.e. the physical
mass of the unbroken symmetry phase), that is to say the solution of the
mass--gap eq.~(\ref{Massgapeq}), generalized to $N>1$, in zero field.
\begin{equation}
\label{massrelation}
       m^2= m_0^2 + \tfrac16 ( N+2 ) \lambda_0 \,I_E^{(1)} ( m^2, \Lambda ) 
\end{equation}  
Let us then define the free mode functions $u_{\bm{k}\,a}^{(0)}$ as solutions of 
\begin{equation} \label{kleingordon}
  ( \Box + m^2 )u_{\bm{k}\,a}^{(0)} ( x ) = 0
\end{equation}
with the same initial conditions of the exact mode functions, that is 
\begin{equation} \label{inifmf}
  u_{\bm{k}\,a}^{(0)}(\bm{x}, 0) = u_{\bm{k}\,a}(\bm{x}, 0)\quad,\qquad
  \dot u_{\bm{k}\,a}^{(0)}(\bm{x}, 0) = \dot u_{\bm{k}\,a}(\bm{x}, 0)
\end{equation}
We need also the free retarded and advanced Green functions
\begin{equation*}
  \begin{split}
    G_{R,ij}^{(0)} ( x - y ) &= \int \frac{d^4 p }{(2 \pi)^4}
    \frac{\delta_{ij}}{ p^2 - m^2 + i \epsilon p_0 }~ e^{-i p ( x - y
      )} \\ G_{A,ij}^{(0)} ( x - y ) &= \int \frac{d^4 p }{(2
      \pi)^4}~ \frac{\delta_{ij}}{ p^2 - m^2 -i \epsilon p_0 }~ e^{-i p
      ( x - y )} 
  \end{split}
\end{equation*}
which allow to cast eq.~(\ref{HFeqsmf}) in the form of an integral equation
(this is a quite standard way to proceed in the renormalization of
out--of--equilibrium problems, see i.e. \cite{Baacke:1996se})
\begin{equation} \label{integrequation}
  u_{\bm{k}\,a,\,i}(x) = u_{\bm{k}\,a,\,i}^{(0)}(x) + \int
  d^4y\,G^{(0)}_{R~in} (x-y)\,{\cal V}_{nj}(y) \, u_{\bm{k}\,a,\,j}(y)
\end{equation}
where
\begin{equation}\label{Omegadef}
  {\cal V}_{ij}(x) = \tfrac13\lambda_0\, \xi_{ij}( x ) + 2\,
  \F'[\xi]_{ij} ( x ) - m^2 \delta_{ij}
\end{equation}
plays evidently the role of mean field for the mode functions. In these
equations, and everywhere else from now on, all field ($\phi$, mode
functions, $\F'$, ${\cal V}$, etc...) are to be thought as defined only for
positive times (initial conditions are at the limit point $t=0^+$) and all
time integrations are restricted to positive values, as appropriate in an
initial value problem.

In compact notation eqs.~(\ref{integrequation}),~(\ref{Omegadef}) can be
written as
\begin{equation} \label{compactnot}
  u_{\bm{k}\,a} = u_{\bm{k}\,a}^{(0)} + G_R^{(0)} {\cal V}\,
  u_{\bm{k}\,a} \quad, \qquad
  {\cal V} = \tfrac13\lambda_0 \, \xi  + 2\,\F' - m^2
\end{equation}
where sums over internal $O(N)$ indices and integrations over spacetime
coordinates are understood. In particular, notice that ${\cal V}$, $\xi$ and
$\F'[\xi]$ act as multiplication operators over spacetime.

The solution of eq.~(\ref{compactnot}) formally reads 
\begin{equation*}
  u_{\bm{k}\,a} = [1- G_R^{(0)} {\cal V}]^{-1}u_{\bm{k}\,a}^{(0)} 
\end{equation*}
so that the cutoffed correlation 
\begin{equation*}
  G_{ij}(x,y) = \mathrm{Re} \int_{|\bm p|<\Lambda} \frac{d^3 p }{(2 \pi)^3}
  \, u_{\bm{p}\,a,\,i}(x)\, \overline{u}_{\bm{p}\,a,\,j}(y)
\end{equation*}
can be written
\begin{equation}\label{GOmega}
  G = G[{\cal V}] = [1-G_R^{(0)}{\cal V}]^{-1}\,G^{(0)}\,
  [1-{\cal V}^T G_A^{(0)}]^{-1}
\end{equation}
where
\begin{equation*}
  G^{(0)}_{ij}( x - y )=  \mathrm{Re} \int_{|\bm p|<\Lambda}
  \frac{d^3 p }{(2 \pi)^3}\,u_{\bm{p}\,a,\,i}^{(0)}(x)\,
  \overline{u}_{\bm{p}\,a,\,j}^{(0)}(y)
\end{equation*}
is the free correlation function, which is entirely fixed by the initial
conditions. In conclusion, by substituting, the form
(\ref{GOmega}) in eq.~(\ref{Gammaformula}), we obtain the
sought self-consistent equation for $\F'$ as
\begin{equation}\label{gammaeq}
  \begin{split}
    \F'[\xi]_{ij} &= \tfrac12 \left(m^2 \delta_{ij} 
      + {\cal V}[\xi]_{ij} \right) - \tfrac16\lambda_0\, \xi_{ij}\\[2mm]
    {\cal V}_{ij} &= \tfrac12 \lambda_0 \,\tau_{ijkn}
    \left(\xi_{kn} + I[{\cal V}]_{kn} - I_E^{(1)} \delta_{kn}\right)
  \end{split}
\end{equation}
where we have renamed the correlation at coincident points as
\begin{equation*}
  I[{\cal V}]_{ij}(x) = G[{\cal V}]_{ij}(x,x)
\end{equation*}
Now, following closely the previous subsection, we may use these expressions,
together with the general formula eq.~(\ref{2Eptfunct}), to evaluate the
renormalization conditions. To begin with, we consider the HF vacuum as initial
state for the quantum fluctuations. That is to say that we start from
equilibrium initial conditions for the mode functions by choosing as $t=0$
kernels the solutions of the static problem of the previous subsection
(generalization to $N>1$ case is straightforward), that is 
\begin{equation}\label{initcond}
  \tilde{\G}_{ij} (\bm{k} ) = \frac{ \delta_{ij}}{2 \omega ( \bm{k} )}\;,
  \quad \omega ( \bm{k} ) = \sqrt{ \bm{k}^2 + m^2} \quad,\qquad 
  \tilde{\s}_{ij} ( \bm{k} )=0
\end{equation}
Let us stress that, in spite of this choice, we are still considering
an out--of--equilibrium problem since we allow for generic initial
conditions for the background field. We will discuss the use of different
initial kernels later on.
  
By the standard prescription we have to compute, using
eq.~(\ref{2Eptfunct}) and eq.~(\ref{gammaeq}), (the Fourier transform
of) the two-- and four--legs vertex functions $\Gamma^{(2)}$ and
$\Gamma^{(4)}$ at some special values of the external momenta
characterized by some given scale $s$.  Owing to our assumption of
unbroken $O(N)$ symmetry, we may choose to renormalize $\Gamma^{(2)}$
at zero momentum, while for $\Gamma^{(4)}$ we make the usual choice of
the symmetric configuration. These special values of $\Gamma^{(2)}$
and $\Gamma^{(4)}$ will be identified, respectively, with the physical
squared mass at equilibrium and with the coupling constant
renormalized at the scale $s$.

Using the relations (\ref{mfinival}) and definitions (\ref{kleingordon}),
(\ref{inifmf}) we obtain the following form for the free mode functions
\begin{equation*}
  u^{(0)}_{\bm{k}\,a,\,i}(x) = \frac{\delta_{ia}}{\sqrt{2
      \omega ( \bm{k} )}}~ e^{-i \omega ( \bm{k} ) t + i \bm{x}
    \bm{k}}
\end{equation*}
The two--legs function is obtained by calculating (\ref{gammaeq}) in zero $\xi$
field. It is easy to check that ${\cal V} = 0$ is solution of eq~(\ref{gammaeq})
for $\xi = 0 $ if the mass renormalization eq.~(\ref{massrelation}) holds true.
Using this fact and the general definition of vertex functions,
eq.~(\ref{2Eptfunct}), we obtain the free two--points vertex function
\begin{equation*}                                    
        \Gamma_{ij}^{(2)} ( x| y ) =(\Box + m^2 )~\delta_{ij}~\delta^4(x-y)
\end{equation*}  
We recall that this is a vertex function in the physical
representation of the CTP formalism, with one $\phi_\Delta$ and one
$\phi_c$ insertions.

Now we have to calculate the four--point function (one $\phi_\Delta$ and
three $\phi_c$). From the variation of the second eq.~(\ref{gammaeq}) with
respect to $\xi$ at $\xi=0$ we find the integral equation
\begin{equation}\label{Thetaeq}
  \vartheta_{ijkm}(x) = \lambda_0\,\tau_{ijkm}\,\delta^{(4)}(x) - \tfrac12
  \lambda_0\,\tau_{ijrs} \int d^4 x' \,I^{(2)}(x-x')\,\vartheta_{rskm}(x')
\end{equation}
for the function
\begin{equation*}
  \vartheta_{ijrs}(x-x') = 2 \frac{\delta {\cal V} \left[ \xi \right]_{ij} ( x )}
  {\delta \xi_{rs} ( x' ) } \Big|_{\xi=0}
\end{equation*}
where
\begin{equation*}
   I^{(2)}(x) = -\frac2{N}\, G^{(0)}_{R,ij}(x)\, G^{(0)}_{ij}(x) 
\end{equation*}
is the (suitably normalized) loop with one retarded and one correlation
Green function.  Eq.~(\ref{Thetaeq}) becomes algebraic in Fourier space
\begin{equation}\label{fThetaeq}
  \tilde\vartheta_{ijkm} ( p )= \lambda_0\tau_{ijkm} -  \tfrac12\lambda_0
  \,\tilde I^{(2)}(p) \,\tau_{ijrs}\,\tilde\vartheta_{rskm}(p)
\end{equation}  
and is easily solved by
\begin{equation}\label{thetaexpl}
  \tilde\vartheta_{ijkm} ( p\,;\lambda_0 )= \tfrac13\,
    g_1(p\,;\lambda_0)\,\big[\delta_{ik}\delta_{jm} + \delta_{jm}\,\delta_{ik}
    \big] + \tfrac13 \, g_2 (p\,;\lambda_0)\, \delta_{ij}\,\delta_{km}
\end{equation}
with $g_1$ and $g_2$ defined as
\begin{equation}\label{gcouplings}
    g_1 (p\,;\lambda_0 ) = \frac{\lambda_0}{1 + \frac13 \lambda_0\, 
      \tilde I^{(2)} ( p )} \quad,\qquad 
    g_2 (p\,;\lambda_0 )= \frac{g_1 (p\,;\lambda_0)} {
      1 + \frac16( N + 2 )\lambda_0\, \tilde I^{(2)} ( p )}
\end{equation}
We recall that, as appropriate to a causal initial value problem, all
Fourier transforms are analytic in the upper complex $p_0$--halfplane.
Notice also that $\tilde I^{(2)}(p) = \tilde I^{(2)}(p,m^2,\Lambda)$ depends 
on the UV cutoff $\Lambda$ through the initial time correlation function
$G^{(0)}_{ij}(x)$,
\begin{equation}\label{I2int}
  \tilde I^{(2)} ( p, m^2, \Lambda ) = -\frac{2}{N}\int_{|\bm k|<\Lambda} 
  \frac{d^4 k} {(2 \pi)^4} \,\tilde G^{(0)}_{R~ij}(p-k)\,\tilde G^{(0)}_{ij}(k)
\end{equation}
so that, as long as $\Lambda$ is finite, it is a function of both the time
component $p_0$ and of the Euclidean 3D length $|\bm p|$, rather then just
of the Lorentz invariant $p^2=p_0^2-{\bm p}^2$.

We now apply the relation between ${\cal V}$ and $\F'$
in~(\ref{integrequation}) and the general definition~(\ref{2Eptfunct}) and
obtain the following form of the four--point vertex function
\begin{equation}\label{4PTsfunct}
  \begin{split} 
    &\Gamma^{(4)}_{i|jkm}(x_1|x_2,x_3,x_4)=\int \Big[\prod_{n=1}^3
    \frac{d^4 p_n }{(2 \pi)^4}\Big] \hat{\Gamma}^{(4)}_{i|jkm}(p_2,p_3,p_4)\, 
    e^{-i\sum_n p_n\cdot (x_1-x_n)}\\[2mm]
    &\tilde{\Gamma}^{(4)}_{i|jkm}= -2 \lambda_0\, \tau_{ijkm} +
      \tilde\vartheta_{ijkm} (p_3+p_4) + \tilde\vartheta_{ikjm}(p_2+p_4) + 
      \tilde\vartheta_{imkj} (p_2+p_3)
  \end{split}
\end{equation}
Notice the structure of $\tilde{\Gamma}^{(4)}$ as sum of three terms, for
the $s$, $t$ and $u$ channels, with the same functional form given by
$\tilde\vartheta_{ijkm} (p)-2\lambda_0/3$. This structure is characteristic
of the mean field approximation.

In the present out--of--equilibrium context, we can define the symmetric
point at the scale $s$ as
\begin{equation*}
  p_3+p_4 = p_3+p_2 = p_2+p_4 = q_s \equiv (0,\bm q)\;,\quad |\bm q|=s
\end{equation*}
with an arbitrary direction ${\hat{\bm q}}$. Evaluating $\tilde{\Gamma}^{(4)}$
at this point at yields the renormalization condition
\begin{equation}\label{rencondcc}
  \lambda \,  \tau_{ijkm} = \tilde{\Gamma}^{(4)}_{i|jkm}
  \big|_{{\rm sym. pt.}\mu} \Longrightarrow \;  \lambda = 
  -2 \lambda_0 + 2 g_1 (q_s\,;\lambda_0) + g_2 (q_s\,;\lambda_0) 
\end{equation}
This result should be compared to the corresponding one in the static $N=1$
case, that is eqs.~(\ref{barepar}),~(\ref{gdef}). We see that, in the
current context with generic $N$ and generic renormalization scale, the
effective coupling $3g$ is split into $2g_1$ and $g_2$, both
scale--dependent.

Before going any further in discussing renormalizability and RG invariance
(independence on the scale $s$), it is convenient to calculate a
``quasi-renormalized'' form of the self--consistent eq.~(\ref{gammaeq}) for
$\F'$ analogous to the eq.~(\ref{eqpotQR2}) found for the effective
potential.

First of all let us define $J[{\cal V}]$ as the part of the integral term
$I[{\cal V}]$ in eq.~(\ref{gammaeq}) that doesn't contain any divergent
integral
\begin{equation*}
  J(x) = I(x) - \left[G^{(0)} +G_R^{(0)}\,{\cal V}\,G^{(0)}
  + G^{(0)}\, {\cal V}^T \, G_A^{(0)}\right](x,x)
\end{equation*}
that is, in Fourier space
\begin{equation}\label{Jfourier}
  \tilde{J}_{km} ( p ) = \tilde{I}_{km}(p) - I_E^{(1)} \delta^{(4)}(p)\delta_{km} +
  \tilde I^{(2)}(p)\, \tilde{{\cal V}}_{km}(p)
\end{equation}
where $\delta^{(4)}(p)$ stands for $-i(p_0+i\epsilon)^{-1}\delta^{(3)}(\bm p)$,
as appropriate in this out--of--equilibrium context.

The second of eqs.~(\ref{gammaeq}) can now be rewritten as
\begin{equation}\label{QRFeq1}
  \tilde{{\cal V}}_{ij} ( p ) = \tfrac12 \tilde\vartheta_{ijkm}
  (p\,;\lambda_0) \left\{ \hat{\xi}_{km} ( p ) + 
    \tilde{J}[ {\cal V}]_{km} ( p )\right\}
\end{equation}
Now, we can define
\begin{equation}\label{DeltaGamma} 
  \Delta \tilde\F'_{ij}( p ) = \tilde\F'_{ij} ( p ) -
  \tfrac12 m^2 \delta_{ij} \delta^{(4)} ( p )- \left[\tfrac14
    \tilde\vartheta_{ijkm}(p)- \tfrac16 \lambda_0\delta_{ik} \delta_{jm} 
  \right] \tilde{\xi}_{km} ( p )
\end{equation} 
and, by substituting in~(\ref{QRFeq1}), we obtain the HF self--consistency
equation in the ``quasi-renormalized'' form that we were looking for, that is
\begin{equation}\label{QRF} 
  \begin{split} 
    \Delta  \tilde\F'_{ij}( p ) = \tfrac14 \tilde \vartheta_{ijkm} ( p )\, 
    \tilde{J} [ {\cal V} ]_{km} (p)\\[1.8mm] 
    \tilde{{\cal V}}_{ij} (p)= 2 \Delta \tilde\F'_{ij} ( p )+ 
    \tfrac12\tilde\vartheta_{ijkm} ( p )\,\tilde{\xi}_{km} ( p )
  \end{split}
\end{equation} 
which is to be compared with eq.~(\ref{eqpotQR2}).

We can now repeat the analysis of the previous subsection in
this more general case. As we will see, conclusions will be rather
similar.

First of all, exactly as it happened in the static case, the
renormalization of the mass parameter receives contributions form
daisy and superdaisy tadpoles with the topologies shown
in~(\ref{massren}). These contribution are actually not affected by
momenta entering in the two--legs function and simply define the bare
mass in terms of the renormalized one according to
eq.~(\ref{massrelation}). From now on we deal with dressed propagators,
i.e. propagators with the renormalized mass.

We look then at the coupling renormalization in eq.~(\ref{rencondcc}).  It
exhibits the same pathological behaviour of the static $N=1$ case,
eq.~(\ref{barepar}), now with a $N-$dependent ``Landau obstruction''.  Most
importantly, from the explicit form of the HF four--legs function in
eq.~(\ref{4PTsfunct}) we see that imposing a finite value to $\lambda$ at a
certain chosen scale $s$ fails to render finite the four--legs function
with generic incoming momenta and, in particular, the coupling $\lambda'$
at any other scale $s'\neq s$. We can correct both shortcomings with the
same strategy of the static $N=1$ case. In fact, the diagrammatic analysis
of the HF four--legs function shows that it is given by the resummation of
``bubble'' graphs
\begin{equation}
\label{ChaindiagrPR}
  \centering
  \includegraphics[width=0.65\textwidth]{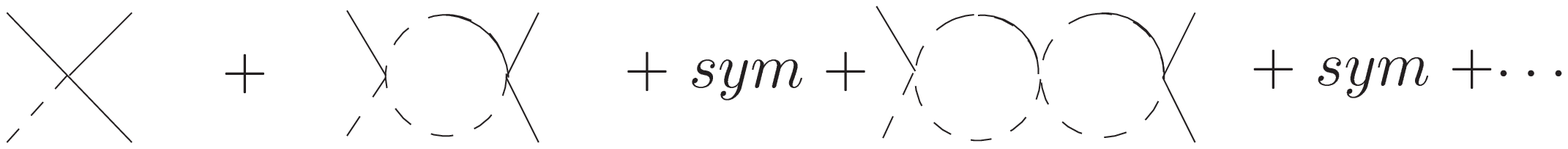}
\end{equation}
where we recall that dotted (solid) lines represent $\phi_\Delta$
($\phi_c$) propagators and external legs coupled with the vertices
(\ref{PRvertices}) of the physical representation. This resummation misses
all leading logarithmically divergent contributions coming from non bubble
diagrams.  In the usual in--out formalism their inclusion provides the
standard 1LRG relation
\begin{equation*}
  \frac1\lambda \simeq \frac1\lambda_0 + \beta_0\lambda\,
  \log\Lambda + \ldots
\end{equation*}
where $\beta_0=\tfrac{N+8}{48\pi^2}$ is the first coefficient of the beta
function and the dots stand for the scheme--dependent finite parts.  Now
the HF approximation dictates a precise scheme where leading logarithms and
finite parts must all be contained in the 1-loop bubble $\tilde
I^{(2)}(q_s,m^2,\Lambda)$, implying the complete parametrization
\begin{equation}\label{RGIcouplingN}
  \lambda_0(\lambda, s/m, \Lambda/m)\big|_{\rm 1LRG} =
  \frac{\lambda}{1 - \tfrac{N+8}{6} \,
    \lambda\,\tilde I^{(2)}(q_s,m^2,\Lambda)}
\end{equation}
which generalizes to the present non-static case with generic $N$ the static
$N=1$ result in eq.~(\ref{RGIcoupling}) and replaces the pure HF parametrization
of eq.~(\ref{rencondcc}). Notice that in the out--of--equilibrium formalism
there could be a priori a complication with non--local finite parts, since the
four--legs function $\tilde{\Gamma}^{(4)}$, even if evaluated at the symmetric
point, need not be in principle completely symmetric in its four $O(N)$ indices
(recall that it has one $\phi_\Delta$ leg and three $\phi_c$ legs). However the
symmetry in the three $\phi_c$ legs is enough to fix the $\tau$ tensor as the
unique $O(N)-$invariant fourth--rank tensor, so that a unique coupling constant
renormalization can occur, including finite parts, just like in the standard
in--out formalism.

We need now to determine how the new contributions which change the HF
parametrization of eq.~(\ref{rencondcc}) to the new one in
eq.~(\ref{RGIcouplingN}) modify the four--legs function away from the
symmetric points of the external momenta. Consistently with the mean field
nature of the approximation, $\tilde{\Gamma}^{(4)}$ must be the sum of
three separated channels with the same functional form $\lambda_0
f[\lambda_0 \tilde I^{(2)} (p)]$ valid at the symmetric point. Then the
request of finiteness plus agreement with the perturbative one-loop result
uniquely fixes the following specific form
\begin{equation}\label{fourpoints}
        \tilde{\Gamma}^{(4)}_{i|jkm} (p_2,p_3,p_4)= 
        \lambda_r(p_3 + p_4)\, \ell_{ikjm} + \lambda_r(p_2 +
        p_4)\, \ell_{ijkm}+\lambda_r(p_3 + p_2)\, \ell_{imjk}
\end{equation}
where the tensor $\ell$ reads
\begin{equation*}
  \ell_{ijkm}= \frac{1}{3(N+8)} \Big[ (N+4)
    \delta_{ij}\delta_{km} + 2 (\delta_{ik} \delta_{jm} + \delta_{im}
    \delta_{jk} ) \Big]
\end{equation*}
and $\lambda_r(p)$ is the running coupling constant, connected to
coupling $\lambda=\lambda_r(q_s)$ at the scale $\mu$ by the relation
\begin{equation}\label{runningcc}
  \frac1{\lambda_r(p)} = \frac1\lambda \,+\, \frac{N+8}6 
    \big[\tilde I^{(2)}(p) - \tilde I^{(2)}(q_s) \big]
\end{equation}
that involves by construction only a subtracted integral, so that one may
now take the limit $\Lambda\to\infty$. Of course, this defines a meaningful
$\lambda_r(p)$ only for $p$ timelike or $p$ spacelike but such that $-p^2$ is
smaller than the Landau pole.

At any rate, eq.~(\ref{fourpoints}) shows explicitly that requiring a
finite coupling strength at a scale $s$ makes the whole four--legs function
finite. Furthermore, the expression~($\ref{runningcc}$) is by construction
RG invariant. Therefore it seems sensible to require such a form for the
coupling and modify the HF resummation by including the missing
contributions that appear in the CTP diagrams with the topology in
eq.~(\ref{RGIcoupling}).

We still have to consider the higher order vertex functions. These are
encoded by construction in the functional $ \Delta \tilde\F'_{ij}$ defined
in~(\ref{DeltaGamma}). From the HF self--consistency equation~(\ref{QRF})
for $ \Delta \tilde\F'_{ij}$ we can see that an explicit dependence on the
logarithm of the cut-off is still present even after the coupling constant
renormalization. The reason is in the presence of the effective vertex
$\vartheta_{ijkm}(p)$ in eq.~(\ref{QRFeq1}), which corresponds to CTP
diagrams of the same type of those drawn in~(\ref{gcoupling}).  Exactly in
the same way of the $N=1$ static case [see eqs.~(\ref{higerorder})
and~(\ref{gcoupling})] we have to include other contributions containing the
correct leading logarithms that come from the $\phi_\Delta-\phi_c$ diagrams
corresponding to those in eq.~(\ref{missdiagr1}) and~(\ref{missdiagr2}). In
order to obtain renormalizability and maintain the single channel structure
of the effective vertex (which is a direct consequence of the mean--field
approximation), these contribution have to be taken momentum dependent in
such a way to determine a Leading Log structure for the effective vertex in
place of $\vartheta_{ijkm}(p)$. Then finiteness and agreement with the lowest
pertubative order require the resulting structure to be simply $\lambda_r(p)\,
\tau_{ijkm}$.

In conclusion, we are naturally led to consider an extended resummation
of diagrams that will satisfy the requests of renormalizability and RG
invariance, by modifying eq.~(\ref{QRF}) in the following way
\begin{equation}\label{modifHF}
  \begin{split} &\tilde\F'_{ij}( p ) = \Delta \tilde\F'_{ij} ( p ) +
  \tfrac12 m^2 \delta_{ij} \delta^{(4)} ( p )+ \tfrac14 \lambda_r
  (p) \, \ell_{ijkm} \, \tilde{\xi}_{km} ( p )\\ &\Delta \tilde\F'_{ij}(
  p ) = \tfrac14 \tilde \lambda_r (p) \, \tau_{ijkm} \, \tilde{J} [
  {\cal V} ]_{km} (p)\\[1.8mm] & \tilde{{\cal V}}_{ij} (p)= 2 \Delta
  \tilde\F'_{ij} ( p )+ \tfrac12 \lambda_r (p)
  \tau_{ijkm}\,\tilde{\xi}_{km} ( p ) \end{split}
\end{equation} 
The renormalized equations of motion are then
\begin{equation} \label{finaleqs}
  \begin{split} &\left[( \Box +m^2)\delta_{ij} + {\cal V}_{ij}
      (x)- \tfrac13 \Xi_{ij}(x) \right]\phi_j ( x )=0 \\[1.8mm]
    & \left[ (  \Box + m^2 ) \delta_{ij} + {\cal V}_{ij} (x) \right]
    u_{\bm{k}\,a,\,j}(x)=0\\[1.8mm] 
    &\tilde{{\cal V}}_{ij}(p)= \tfrac12 \lambda_r(p)\,\tau_{ijkm}
    \left[\tilde{\xi}_{km}(p) + \tilde{I}_{km}(p) - I_E^{(1)} 
      \delta^{(4)}(p)\delta_{km} +
      \tilde I^{(2)}(p)\, \tilde{{\cal V}}_{km}(p)\right]
  \end{split}
\end{equation}
where $\Xi_{ij}(x)$ is the inverse Fourier transform of $\tilde\Xi_{ij}(p) =
\tfrac32 \lambda_r(p)\,(\tau_{ijkm}-\ell_{ijkm})\,\tilde{\xi}_{km}(p)$ and we
recall that $\xi(x)=\phi_i(x)\phi_j(x)$ and $\tilde{I}_{km}(p)$ is the Fourier
transform of
\begin{equation*}
  I[{\cal V}]_{ij}(x)= \mathrm{Re} \int_{|\bm p|<\Lambda} \frac{d^3 p
  }{(2 \pi )^3}~ u_{\bm{p}\,b,\,i}(x)\, \bar u_{\bm{p}\,b,\,j}(x)
\end{equation*}
which is a functional of the mean field ${\cal V}$ through the evolution
equation for the mode functions.
 
The absence of any divergent cut--off dependence in the modified HF
approximation defined by eqs.~(\ref{modifHF}) or by the equations of motion
(\ref{finaleqs}) is manifest, since they involve only the finite running
coupling constant and the fully subtracted (Fourier transform of) the equal
point correlation $I_{ij}(x)$. Comparing the expression for $\tilde{\cal
  V}_{ij}(p)$ with the original one in eq.~(\ref{gammaeq}), that reads in
Fourier space,
\begin{equation*}
    \tilde{\cal V}_{ij}(p) = \tfrac12 \lambda_0 \,\tau_{ijkn}
    \left[\tilde\xi_{kn}(p) + \tilde I_{kn}(p) - 
      I_E^{(1)}\delta^{(4)}(p)\delta_{kn}\right]
\end{equation*}
we see that our improved renormalization simply amounts to perform the
logarithmic subtraction on $I[{\cal V}]_{ij}$ while promoting the bare coupling
$\lambda_0$ to the running one $\lambda_r(p)$. As a consequence, the
$p$--dependence of $\lambda_r(p)$ and $\tilde I^{(2)}(p)$ imply a space--time
nonlocality of the modified HF self--consistency equation for the mean field
${\cal V}(x)$, which now depends on the whole history of the background $\phi$
and of the mode functions $u_{\bm{k}\,a}$ from time zero to time $t$. A similar
space--time nonlocality appears in the equation of motion for the background
field $\phi$. Causality in this nonlocal evolution is guaranteed by analyticity
in the upper $p_0-$halfplane of $\tilde I^{(2)}(p)$.  Let us observe, however,
that the third equation in eqs.~(\ref{finaleqs}) does not provide an explicit
expression of the mean field ${\cal V}$ in terms of the background and of the
mode functions. To obtain such an expression, we must solve for ${\cal V}$,
paying the price of losing manifest finiteness. In practice, this reduces to the
simple matrix inversion already performed on eq.~(\ref{Thetaeq}) and yields
\begin{equation*}
    \tilde{\cal V}_{ij}(p) = -\tfrac12 \,\tilde\vartheta_{ijkm}
    (p\,;-\lambda_r(p)) \big[\tilde{\xi}_{km}(p) + \tilde{I}_{km}(p) - 
    I_E^{(1)}  \delta^{(4)}(p)\delta_{km} \big] 
\end{equation*}
where the $\vartheta_{ijkm}(p\,;u)$ is the function defined in
eq.~(\ref{thetaexpl}). In coordinate space this becomes
\begin{equation}\label{Vexplicit}
  \begin{split}
    {\cal V}_{ij}(x) = \frac13 \int d^4x'\, &\Big\{
    2\,\gamma_1(x-x') \Big[\phi_i(x')\phi_j(x') + I_{ij}(x') -
    \delta_{ij}I_E^{(1)} \Big] \\    &+ \;\delta_{ij}\,\gamma_2(x-x') 
    \Big[ \phi^2(x') + I_{kk}(x') - NI_E^{(1)} \Big]\Big\}
  \end{split}
\end{equation}
where $\gamma_1(x)$ and $\gamma_2(x)$ are the inverse Fourier transforms of the
functions $-g_1(p\,;-\lambda_r(p))$ and $-g_2(p\,;-\lambda_r(p))$, respectively,
which can be read from eq.~(\ref{gcouplings}) and are indeed analytic in the
upper $p_0-$halfplane. For instance, we have explicitly
\begin{equation*}
  \gamma_1(x) = \int \frac{d^4 p }{(2\pi)^4} \frac{\lambda \; 
    e^{-i p \cdot x}}{1+ \frac16\lambda \big[(N+1)\tilde I^{(2)}(p) - 
    (N+\tfrac43)\tilde I^{(2)}(q_s) \big] }
\end{equation*}
with an even more involved expression for $\gamma_2(x)$.  Clearly $\gamma_{1,2}$
depend logarithmically on the cut--off $\Lambda$ through an incompletely
subtracted $\tilde I^{(2)}(p)$, but, by construction, this logarithms cancels
out with those coming from the incomplete subtraction $I_{ij}(x) -
\delta_{ij}I_E^{(1)}$ on the equal point correlation of the mode functions.
The memory functions $\gamma_1(x)$ and $\gamma_2(x)$, as well as
the inverse Fourier transform of $\lambda_r(p)$ which enters in the
evolution equation of the background field [see eq.~(\ref{finaleqs})], all
share the same behaviour for large timelike values of $x$, namely 
a $(x^2)^{-1/2}$ decay modulated by oscillating factors.

When $N=1$ things simplify considerably and we find for the mean field
\begin{equation*}
  {\cal V}(x) = \int d^4x'\, \gamma(x-x') \Big\{ \phi^2(x') + 
  \int\frac{d^3 k}{(2 \pi )^3}\Big[ 
  |u_{\bm k}(x)|^2 - \frac1{2\omega(\bm k)}\Big] \Big\} 
\end{equation*}
where 
\begin{equation*}
  \gamma(x) = \tfrac23 \gamma_1(x) + \tfrac13 \gamma_2(x) =
  \tfrac12\int \frac{d^4 p }{(2\pi)^4}\frac{\lambda \; e^{-i p \cdot x}}
  {1+ \lambda  \big[I^{(2)}(p) - \tfrac32 I^{(2)}(q_s)\big] }
\end{equation*}
while mode functions and  background evolve according to
\begin{equation*}
  \begin{split} 
    & \left[ \Box + m^2 + {\cal V}(x) \right] u_{\bm k}(x)=0 \\
    \Big[ \Box +m^2 + \;&{\cal V}(x)- \tfrac13 \int d^4x'\, 
    \gamma_r(x-x') \phi^2(x') \Big] \phi( x )=0
  \end{split}
\end{equation*}
where 
\begin{equation*} 
  \gamma_r(x) = \int \frac{d^4 p }{(2\pi)^4}\frac{\lambda \; e^{-ip\cdot x}}
  {1+ \tfrac32 \lambda  \big[I^{(2)}(p) - I^{(2)}(q_s)\big] }
\end{equation*}
is the ultraviolet finite inverse Fourier transform of the running coupling
constant $\lambda_r(p)$.

On the opposite end, when $N\to\infty$, one can verify that the standard
local evolution equations \cite{Cooper:largeN} are recovered from
eqs.~(\ref{finaleqs}). To this end, it is convenient to restrict first to
the special case of a background field which maintains a fixed direction,
that is $\phi_i(x) = \phi (x)v_i$, with $v$ some fixed $N-$dimensional unit
vector. In fact in this case we can reduce eqs.~(\ref{finaleqs}) to an
index-free form by introducing longitudinal and transversal projectors
\begin{equation*}
  P_{L,ij}=v_iv_j\quad, \qquad P_{T,ij}=\delta_{ij} -P_{L,ij}
\end{equation*}
Then we can set
\begin{equation*}
  \begin{split}
    & u_{\bm{k}\,a,\,i}(x)= u_{\bm{k}\,L}(x)P_{L,ia} + 
    u_{\bm{k}\,T}(x)P_{T,ia}\\[1.8mm]
    &{\cal V}_{ij}(x)= {\cal V}_L(x)P_{L,ij} + {\cal V}_T(x)P_{T,ij}
  \end{split}
\end{equation*}             
which are compatible with initial condition in eq.~(\ref{initcond}) if we
assume that the longitudinal and transverse parts of the mode functions and
their derivatives are equal at initial time. By substituting into the
equations of motion and projecting one obtains
\begin{equation} \label{EqsTL}
  \begin{split} 
    &[\Box + m^2 + {\cal V}_T (x)] u_{\bm{k}\,T}(x) =0 \\ 
    &[\Box + m^2 + {\cal V}_L (x)] u_{\bm{k}\,L}(x) =0 \\ 
    &[\Box + m^2 +\tilde{\cal V}_L (x)- \tfrac13 \Phi (x)] \phi(x)=0\\
    &\tilde{\cal V}_L (p) = \tfrac12
    \tilde{\Phi} (p) + \tfrac16 \lambda_r(p) \Big[(N-1) \tilde J_T(p)
    + 3  \tilde J_L(p) \Big]\\&
    \tilde{\cal V}_T(p) = \tfrac16
    \tilde{\Phi} (p) + \tfrac16 \lambda_r(p) \Big[(N+1)\tilde J_T(p)
    + \tilde J_L(p) \Big]
  \end{split}
\end{equation}  
where $\Phi(x)$ is the inverse Fourier transform of $\tilde{\Phi}(p)=
\lambda_r(p)\,\tilde{\xi}(p)$ [recall that $\xi(x)= \phi^2(x)$] and the meaning
of $J_L$ and $J_T$ is obvious [see eq.~(\ref{Jfourier})]. Again, the formulation
in momentum space for the mean field ${\cal V}$, while convenient to avoid
lengthy convolution integrals and to allow for an easy check of finiteness, does
not make causality manifest. The opposite situation, with finiteness hidden and
causality manifest would follow by solving explicitly for ${\cal V}$ and then
reverting to coordinate space as done above for the general case [see
eq.~(\ref{Vexplicit})] or for the $N=1$ case. 

Now let us rescale the coupling and the background field as prescribed by the
standard large $N$ procedure
\begin{equation*}
  \lambda \to \lambda/N \qquad \phi (x) \to \sqrt{N} \phi (x) 
\end{equation*}
By substituting in (\ref{EqsTL}) and taking the limit on $N$ we obtain (notice
that the longitudinal mode functions decouple)
\begin{equation*}
 \begin{split} &[ \Box + m^2 + {\cal V}_T (x)] u_{\bm{k}\,T}(x)=0 \\ &[
   \Box + m^2 + {\cal V}_T (x)] \phi(x)=0\\&{\cal V}_T (x) = \tfrac16 \lambda
   \Big\{ \phi^2(x) + \int\frac{d^3 k}{(2 \pi )^3}\Big[ 
   |u_{\bm k}(x)|^2 - \frac1{2\omega(\bm k)}\Big]  
   + I^{(2)}(s){\cal V}_T(x)  \Big\}
   \end{split}
\end{equation*} 
which are the usual, local and renormalized large $N$ evolution equations. We
remark that $\lambda$ is the renormalized coupling at the scale $s$ while the
nonlocal terms (with the running coupling $\lambda(p)$) have all disappeared by
explicit cancellation. This is indeed what was to be expected, on the basis of
the same diagrammatic analysis that lead to our improvement of the HF
approximation, since the nonrenormalizable terms in the standard HF
approach, as well as the contribution we have added to cure the problem, all
come from diagrams that are at least $1/N$ suppressed.

We conclude this lengthy subsection with an explicit check of the
Renormalization--Group invariance of our improved HF approximation, by verifying
that indeed all vertex functions obtained by eqs.~(\ref{modifHF}) are solutions
of Callan-Symanzik equation. These equation states the RG invariance of any
observable $\cal O$ which is a function of coordinates $x_i$ (or momenta $p_i$),
of the scale $s$, the coupling $\lambda$ and the parameter $\sigma= m^2/s^2$,
that is
\begin{equation}\label{CSeq}
  \left[ s \frac{\partial}{\partial s} + \beta(\lambda,\sigma)
    \frac{\partial}{\partial \lambda} + \beta_{\phi^2}(\lambda,\sigma)
    \frac{\partial}{\partial \sigma} \right] {\cal O}( \{ x_i
  \}|s,\lambda,\sigma)=0
\end{equation}   
with $\beta$ and $\beta_{\phi^2}$ functions to be determined. Notice the
absence of the term with the anomalous dimension due to the lack of
field renormalization in HF approximation. By applying eq.~(\ref{CSeq}) to
the two--legs function one obtains $\beta_{\phi^2}= - 2 \sigma$.  Then by
applying to four--legs function one obtains
\begin{equation*}
  \beta(\lambda,\sigma)= - \tfrac16(N+8)\lambda^2 \lim_{\Lambda\to\infty}
  \,s\frac{\partial} {\partial s} 
  I^{(2)} (s,m^2,\Lambda) \Big|_{m^2 = \sigma s^2}
\end{equation*}
Then for $s^2\gg m^2$ we have $\beta \simeq \frac{N+8}{48 \pi^2} \lambda^2$
as expected. Now, is easy to check that all free propagators with dressed
mass satisfy the CS equation since they are functions of $m^2$ alone. Then,
since the generic $n$--legs function is a functional of the propagators and
of the RG-invariant running coupling $\lambda_r(p)$ through the four--legs
function, one immediately concludes that it satisfies eq.~(\ref{CSeq}).


\subsection{Other initial states and nonzero temperatures}\label{initstate}

As stated in the previous subsection, with generic initial conditions on the
background field eqs.~(\ref{finaleqs}) already describe an out--of--equilibrium
problem, in spite of the choice in eq.~(\ref{initcond}) of equilibrium initial
conditions on the quantum fluctuations. It is nonetheless sensible to ask
whether and how we can choose different initial conditions for the mode
functions without spoiling the properties of renormalizability and RG
invariance.

What might happen can be shown by the following example. Let us consider, in the
simple $N=1$ case, an initial state of the same form of the HF vacuum but with a
different mass $M$, which could be for instance the solution of the gap equation
(\ref{Massgapeq}) in the case of an uniform background. This is a frequent
choice in dealing with non--equilibrium problems and has a precise physical
meaning, since it corresponds to the minimization of the quantum fluctuation
part of the HF energy at fixed uniform background. At any rate, as soon as we
assume as initial conditions
\begin{equation*}
  \tilde{\G}_{ij} (\bm{k} ) = \frac{ \delta_{ij}}{2 \Omega ( \bm{k} )}\;,
  \quad \Omega(\bm{k} ) = \sqrt{ \bm{k}^2 + M^2} \quad,\qquad 
  \tilde{\s}_{ij} ( \bm{k} )=0
\end{equation*}
with $M\neq m$, then the free mode functions have more than one frequency
component
\begin{equation*} 
  u^{(0)}_{\bm{k}}(x) = \frac{e^{ i \bm{k} \cdot\bm{x}}}{2 \sqrt{2 \Omega(\bm k)}}
  \left\{ \left[ 1 + \frac{\Omega(\bm k)}{\omega(\bm k)} \right] 
    e^{-i \omega(\bm k) t}+ \left[ 1 - \frac{\Omega(\bm k)}{\omega(\bm k)}
    \right] e^{-i \omega(\bm k) t}\right\}
\end{equation*}
and the free correlation function is no longer translationally invariant in
time. Then the integral term in eqs.~(\ref{finaleqs}) [recall that here $N=1$]
\begin{equation*} 
  \tilde{I}(p) - I_E^{(1)} \delta^{(4)}(p) +I^{(2)}(p)\, \tilde{{\cal V}}(p)
\end{equation*}
in spite of the subtractions still contains the superficially divergent
contribution
\begin{equation}\label{initsing}     
  \int \frac{d^3 k}{(2 \pi)^3}\, \frac{m^2- M^2}{4 \omega(\bm k)^3}
  \cos{2 \omega(\bm k) t} 
\end{equation}
that indeed diverges with the cut--off when $t=0$. 
These initial time singularities have first been discussed in
\cite{Baacke:1997zz} and removed by a Bogoliubov transformation on the
initial state, which in practice amounts to a redefinition of the
initial kernel in such a way that the leading terms of an
high--momentum expansion are the same as at equilibrium
\begin{equation*}     
  \hat{\G} ( \bm{k} ) \sim \frac{1}{2 \sqrt{{\bm k}^2}}+
  \frac{m^2}{4({\bm k}^2)^{3/2}}+ \dots
\end{equation*}
In fact, one may verify that the initial singularity (and any other
divergence as well) are absent for any choice of kernel having the above
large $\bm k$ expansion. A simple interpretation is that the
renormalization procedure ensures finiteness for any initial Gaussian state
belonging to the same Fock space of the HF vacuum. From this simple example
we can extrapolate the generic condition on the short--distance behaviour of
the initial state kernel 
\begin{equation*}     
  \hat{\G} ( {\bm x}, {\bm y}) \simeq \frac{1}{4\pi^2|{\bm x}-{\bm y}|}+
  \frac{m^2}{8\pi^2}\log|{\bm x}-{\bm y}| + \dots
\end{equation*}
that ensures the cancellation of all divergent terms are guaranteed by mass and
coupling constant renormalization. This conclusions can be immediately extended
to the generic $N>1$ case.

Before concluding let us say some words about the case when the initial
state has a nonzero temperature $T$. The formalism for pure state dynamics
introduced in section~\ref{CTP} can be easily generalized to statistical
mixtures defined by Gaussian density matrices.
\begin{equation}\label{gaussdensmat}
\begin{split}
   \rho [\varphi_1, \varphi_2] =& \mathcal{N} \exp\Big\{ i
    \braket{p}{\varphi_1-\varphi_2} - \bra{\Delta
    \varphi_1}\left[\tfrac14 \G^{-1} + i\s\right] \ket{\Delta
    \varphi_1}\\-& \bra{\Delta \varphi_2}\left[\tfrac14 \G^{-1} -
    i\s\right] \ket{\Delta\varphi_2}+ \tfrac12 \bra{\Delta
    \varphi_1}\G^{-\tfrac12}\, \zeta \, \G^{-\tfrac12} 
    \ket{\Delta \varphi_2}\Big\}
\end{split}
\end{equation}
Where $\Delta \varphi_1 \equiv \varphi_i - \phi$. The parameters are
the the background field $\phi$ and its momentum $p$, the symmetric kernels
$\G$, $\s$, $\gamma$ and the antisymmetric $\sigma$ (with $\zeta=
\gamma + 4i \G^{1/2}\, \sigma \, \G^{1/2} $). The normalizability of
the state ($Tr[\rho]=1$) requires $ \G^{-1/2}\, (1-\gamma) \,
\G^{-1/2}$ to be positive. In CTP formalism the generalization
proceeds simply by substituting $ \Psi[\varphi_+]  \overline{\Psi} [
\varphi_-] $ with  $\rho [\varphi_+, \varphi_-]$ in the
path integral of eq.~(\ref{CTPgenfun}). Then HF equations are still as
before while more general initial conditions for the mode functions
are allowed. Let us  consider an initial state with  translationally
invariant kernels and, for simplicity, let us put $N=1$,
$\s=\sigma=0$, then we have
\begin{equation*}
\begin{split} 
    u_{\bm{k}}(\bm{x},0) &= [\tilde{\G}(
    \bm{k})]^{1/2}\,[1-\tilde{\gamma}( \bm{k})\,]^{- 1/2}\, e^{i
    \bm{k} \cdot \bm{x}} \\ \dot{u}_{\bm{k}}(\bm{x},0) &= \left[-
    \tfrac{i}2 \tilde{\G}(\bm{k})^{-1}\,[1-\tilde{\gamma}^2
    (\bm{k})]^{1/2}\right] u_{\bm{k}} (\bm{x}, 0)
\end{split}
\end{equation*}
For vanishing initial background field and momentum, standard equilibrium
solutions are given by
\begin{equation}\label{stdTeq}
  \G({\bm k})=\frac{1}{2 \omega_T(\bm k)} \tanh{ \frac{\omega_T(\bm k)}{T}}
  \;,\quad \omega_T(\bm k) = \sqrt{{\bm k}^2 + m_T^2} \;,\quad 
  \gamma_{\bm k}= \left[ \cosh{ \frac{\omega_T(\bm k)}{T}}\right]^{-1} 
\end{equation}
with $m_T^2$ defined by the gap equation
\begin{equation*}
        m_T^2 - m^2 = - I^{(1)}_E + (m_T^2 - m^2) I^{(2)}(0) + \tfrac12
        \lambda\int \frac{d^3 k}{(2 \pi)^3}\, \frac{1}{2 \omega_T(\bm k)
        }\,\coth{ \frac{\omega_T(\bm k)}{2T}}         
\end{equation*}
By expanding $\omega_T(\bm k)$ around $\omega_0(\bm k)=\omega(\bm k)$ one can
easily check the cancellation of divergent terms.  We can then consider the
out--of--equilibrium problem with the same initial conditions on the mode
functions but arbitrary initial values for the background field and its
velocity. The free mode functions read
\begin{equation*} 
         u^{(0)}_{\bm{k}}(x) = 
        \frac{1}{2} 
        \sqrt{\frac{1}{2\omega_T}
        \coth{\frac{\omega_T}{2T}}}
        \left[ \left( 1 + \frac{\omega_T}{\omega} \right) e^{-i \omega
        t}+ \left( 1 - \frac{\omega_T}{\omega} \right) e^{i \omega t}
        \right] e^{ i \bm{x} \bm{k}}
\end{equation*}
Calculating the free correlation function we easily can check that the mean
field expression in the last of eqs.~(\ref{finaleqs}) is free of
divergences, except for the one at $t=0$, which is given by
eq~(\ref{initsing}) with $M=m_T$. As above, this can be cured  by a
Bogoliubov transformation, which in practice amounts to changing 
the equilibrium solution in eq.~(\ref{stdTeq}) by allowing a suitable
ultraviolet $k$-dependence in the temperature.

One last comment regards RG invariance. We can see that the proof of
the subsection~\ref{general} still holds as long as the free correlation
function depends on the renormalized squared mass alone.


\end{document}